\documentclass{article}

\usepackage{PRIMEarxiv}

\usepackage[utf8]{inputenc} 
\usepackage[T1]{fontenc}    
\usepackage{hyperref}       
\usepackage{url}            
\usepackage{booktabs}       
\usepackage{amsfonts}       
\usepackage{nicefrac}       
\usepackage{microtype}      
\usepackage{lipsum}
\usepackage{mathtools}
\usepackage{amsmath}
\usepackage{bm}
\usepackage{stmaryrd}
\usepackage{natbib}
\usepackage{fancyhdr}       
\usepackage{graphicx}       
\graphicspath{{media/}}     

\graphicspath{{./figures/}}

\usepackage{subcaption}
\usepackage{amssymb}
\usepackage{cleveref}
\usepackage{makecell}
\usepackage{xcolor}
\usepackage{tikz}
\usetikzlibrary{matrix, positioning, fit, arrows.meta, calc, petri,
                positioning,shapes,shadows,graphs, backgrounds}

\DeclareMathOperator*{\argmax}{argmax}
\DeclareMathOperator*{\argmin}{argmin}

\DeclarePairedDelimiterX{\norm}[1]{\lVert}{\rVert}{#1}

\definecolor{ao}{rgb}{0.0, 0.5, 0.0}

\newcommand\T{{\mathcal{T}}}
\newcommand\y{{\bf{y}}}
\newcommand\Y{{\bf{Y}}}

\newcommand\M{{\bf{M}}}

\newcommand\bF{{\bf{F}}}
\newcommand\V{{\bf{V}}}
\newcommand\bS{{\bf{S}}}
\newcommand\bs{{\bf{s}}}

\newcommand\bk{{\bf{k}}}
\newcommand\bXi{{\boldsymbol{\Xi}}}

\newcommand\bTheta{{\boldsymbol{\Theta}}}
\newcommand{\R}{{\ensuremath{\mathbb{R}}}}
\newcommand{\reac}[1]{\xrightarrow{#1}}
\newcommand{\gt}[1]{\textcolor{ao}{#1}}
\newcommand{\ft}[1]{\textcolor{red}{#1}}
\newcommand{\at}[1]{\textcolor{orange}{#1}}

\definecolor{darkblue}{RGB}{0,0,139}
\definecolor{inriaorange}{RGB}{240,126,38}
\definecolor{inriamauve}{RGB}{101,97,169}
\tikzstyle{treeline} = [draw, ->, line width=1.2, thick, darkblue]

\tikzset{%
     every place/.style={draw=inriaorange!70,fill=inriaorange!20,thick,
        minimum size=8mm},
     every transition/.style={draw=inriamauve!50,fill=inriamauve!20,thick,
        minimum size=6mm},
     pre/.style={<-,shorten <=1pt,>=stealth,thick},
     post/.style={->,shorten >=1pt,>=stealth,thick},
     null/.style={-, thick},
     round/.style={rounded corners=5pt},
     fire/.style={transition,fill=yellow},
     posreg/.style={->,shorten >=1pt,>=stealth,very thick,green},
     negreg/.style={-|,shorten >=1pt,>=stealth,very thick,red},
     ampersand replacement=\&,
     invisible/.style={opacity=0},
     visible on/.style={alt={#1{}{invisible}}},
     alt/.code args={<#1>#2#3}{%
        \alt<#1>{\pgfkeysalso{#2}}{\pgfkeysalso{#3}} 
     },
}

\tikzset{
    background shadedraw/.style 2 args={draw=#1, #2},
    background shadedraw/.default={}{top color=white, bottom color=white},
    shadedraw on/.style={alt=#1{}{background shadedraw}},
          }

\definecolor{colora}{rgb}{0.12156862745098039, 0.4666666666666667,0.7058823529411765}
\definecolor{colorb}{rgb}{1.0, 0.4980392156862745, 0.054901960784313725}
\definecolor{colorc}{rgb}{0.17254901960784313, 0.6274509803921569, 0.17254901960784313}
\definecolor{colord}{rgb}{0.8392156862745098, 0.15294117647058825, 0.1568627450980392}
\definecolor{colore}{rgb}{0.5803921568627451, 0.403921568627451, 0.7411764705882353}

\tikzstyle{line} = [draw, ->, line width=.8, very thick]
\tikzstyle{treeline} = [draw, ->, line width=1.2, thick, darkblue]

\title{Reactmine: a statistical search algorithm for inferring chemical reactions from time series data}

\author{Julien Martinelli\\Department of Computer Science\\Aalto University, Espoo, Finland\\ \texttt{julien.martinelli@aalto.fi}\And
Jeremy Grignard\\Institut de recherches Servier\\Suresnes, France\\ \And
Sylvain Soliman\\Inria Saclay, Lifeware Group\\Palaiseau, 91120, France\\ \And
Annabelle Ballesta\\Inserm U900, Institut Curie\\Saint Cloud, France\\MINES ParisTech, CBIO, PSL Research University\\Paris, France\\ \And François Fages\\Inria Saclay, Lifeware Group\\Palaiseau, 91120, France\\ \texttt{francois.fages@inria.fr}}

\begin{document}
\maketitle

\begin{abstract}
 Inferring chemical reaction networks (CRN) from concentration time series
  is a challenge encouraged by
  the growing availability of quantitative temporal
    data at the cellular level. This motivates the design of algorithms to infer the preponderant
    reactions between the molecular species observed in a given biochemical process, and build CRN structure and kinetics models.
    Existing ODE-based inference methods such as SINDy resort to least
    square regression combined with sparsity-enforcing penalization,
    such as Lasso. 
    However, we observe that these methods fail to learn sparse models when the input time series are only available in \emph{wild type}
    conditions, i.e.~without the possibility to play with combinations of zeroes in the initial conditions.
    We present a CRN inference algorithm which
    enforces sparsity by inferring reactions in a sequential fashion within a search tree of bounded depth,
    ranking the inferred reaction candidates 
    according to the variance of their kinetics on their supporting transitions, 
    and re-optimizing the kinetic parameters of the CRN candidates on the whole trace in a final pass. 
    We show that Reactmine succeeds both on simulation data by retrieving hidden CRNs
    where SINDy fails, and on two real datasets, one of fluorescence videomicroscopy of cell cycle and circadian clock markers,
    the other one of biomedical measurements of systemic circadian biomarkers possibly acting on clock gene expression in peripheral organs,
    by inferring preponderant regulations in agreement with previous model-based analyses. The code is available at \href{https://gitlab.inria.fr/julmarti/crninf/}{https://gitlab.inria.fr/julmarti/crninf/} together with introductory notebooks.
\end{abstract}

\section{Introduction}

With the automation of biological experiments and the increase of quality of cell
measurements, automating the building of mechanistic models from data becomes
conceivable and a necessity for many new applications. 
The structure of such models, e.g.\ gene regulatory networks (GRN) or chemical reaction networks (CRN), 
is classically built from an extensive
review and compilation of the literature by the modeler. 
More recently,
efforts have been made to develop model learning algorithms to assist
modelers in order to partly automate the model building process, in particular when time series measurements are  available.

Extensive literature is available in the context of GRN inference or unsupervised learning,
partly motivated by knowledge discovery problems, such as presented in the DREAM series of challenges \citep{SMC07anyas},
or experiment design \citep{KWJRBMKO04nature}.
A GRN consists in a directed graph $G = (W, E)$ of genes and edges $E_{ij}$ between genes, whenever a gene transcription factor $W_i$ binds to the
promoter region of target gene $W_j$.
GRN inference algorithms feature a wide range of machine learning methods,
e.g.\ Decision Trees \citep{HG18sr}, Information Theory \citep{ZMC10bmcb} or
Gaussian Processes \citep{AVI20nc}.

Less work concerns CRN inference, i.e.~the problem of inferring both the structure and kinetics of chemical reactions between some molecular species observed with time series data about their concentrations.
The structure of a CRN can be represented by a bipartite directed graph 
with edges from molecular species vertices to reaction vertices, representing the reactants of a reaction,
and edges from reaction vertices to species representing their product.
Of note, the indegree and outdegree of a reaction node can be above one, which allows for bimolecular
reactions like complexations, e.g.~$A+B \rightarrow C$, or catalyzed transformations, e.g.~$A+B\rightarrow A+C$.
Each reaction of a CRN is given with its kinetics, using reaction rate functions such as mass action law,
Michaelis-Menten or Hill kinetics. 
The rate function of each reaction appears as a term in the ordinary differential equations (ODE) that govern
the time evolution of the products and reactants of the reaction.
Overall, both the difference of structure and the importance of the kinetics make the above GRN inference
methods hardly applicable to CRN inference problems.

CRN inference may thus rely on the ODE semantics of a CRN to apply ODE inference methods from time series data,
such as the state-of-the-art tool SINDy (Sparse Identification of Nonlinear Dynamics,~\cite{BPK16pnas})
with an appropriate library of kinetic functions.
The main assumption is that the dynamics of each variable can be expressed
using only a few functions of the observed variables, without introducing hidden variables,
so that techniques like sparse regression can be used to determine the optimal members of the library for a given problem.
Selecting the ground truth sparse set of predictors is however
a task best achieved provided two hypotheses are satisfied: low correlations
between the true predictors and the spurious ones, and low partial correlations
among the set of true predictors \citep{ZY06jmlr}.
These conditions
can reasonably be met in datasets composed of multiple initial states with various combinations
of absent and present species, possibly obtained by silencing genes of interest (i.e. \emph{knockout} experiments)
or exposure to targeted inhibitors. Such datasets containing time series in multiple conditions indeed allow the different
reactions to be witnessed in an independent manner \citep{CFS17cmsb}. 
However, this is not always possible, and in many situations,
like  in the context of experimental time series
data obtained from protein fluorescence microscopy \cite{FKT14pnas},
one has to work only with traces obtained in a \emph{wild type}
setting in which those hypotheses are not satisfied. 

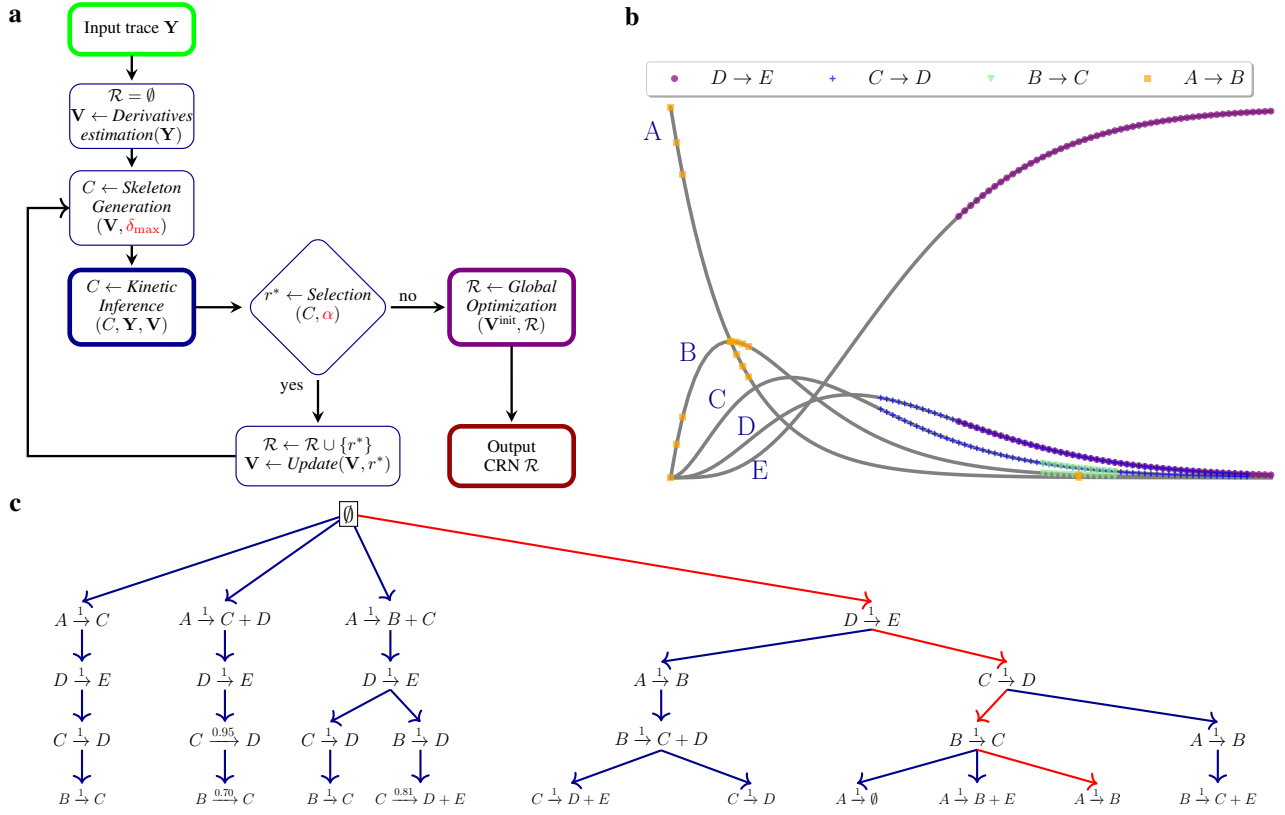
\begin{figure*}[ht!]
\begin{tikzpicture}

 \node[] at (6.7, 4.5) (up)  {\includegraphics[width=.54\linewidth]{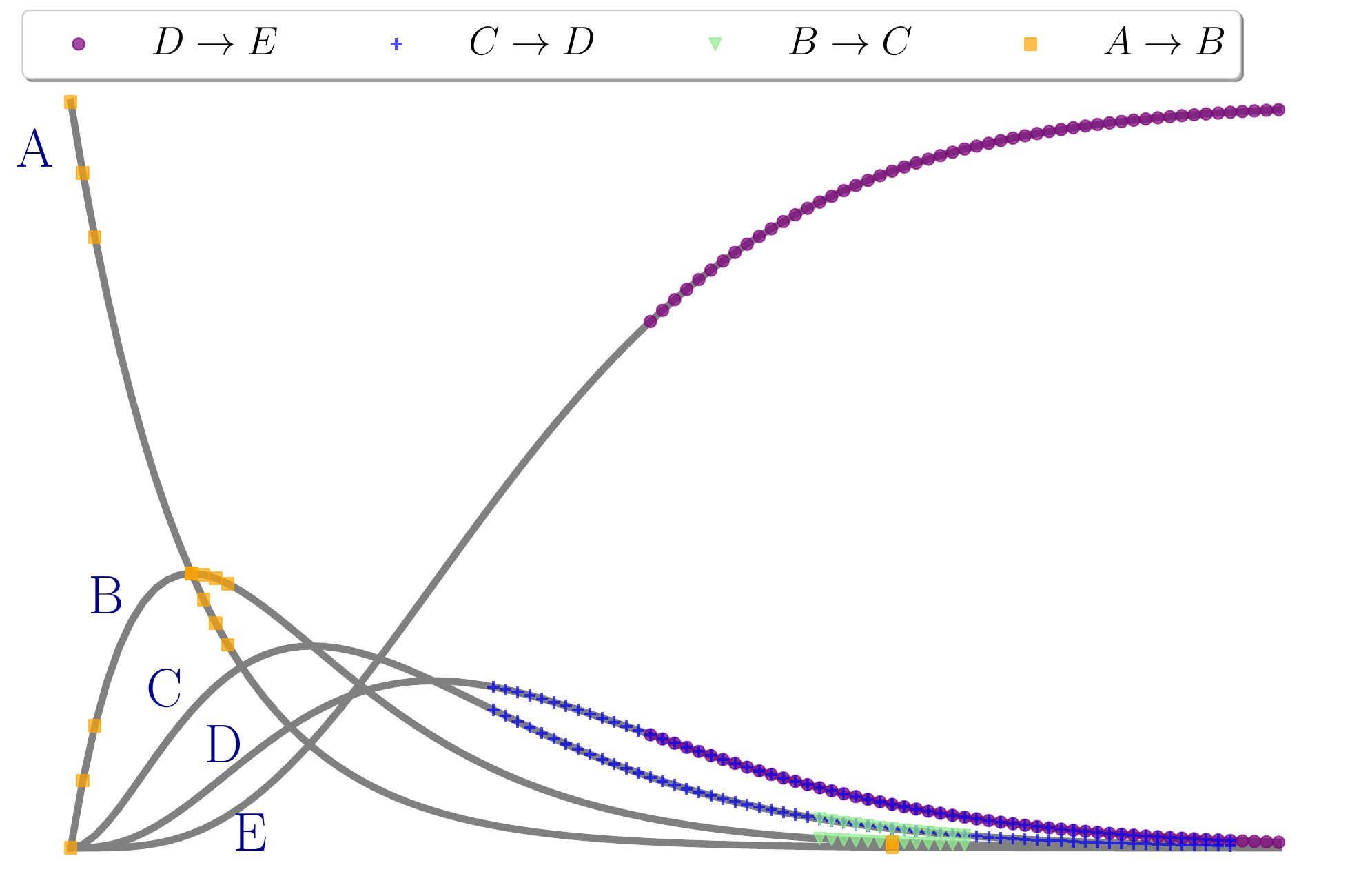}};

\scalebox{1.1}{

\node at (-2,4.5) (algo) {
\begin{tikzpicture}[node distance=1cm, scale=.55,
                    every node/.style={scale=.6},
                    every matrix/.style={nodes={scale=.5}}]

    \node [rectangle, align=center, rounded corners, draw=green,
           ultra thick, inner sep=0pt, minimum height=1cm, minimum
           width=2.5cm]
           (input) {Input trace $\Y$};

    \node [rectangle, align=center, rounded corners,
           inner sep=0pt, minimum height=1.3cm, minimum width=2.5cm,
           below of=input, node distance=1.75cm, draw=darkblue] (velo)
           {$\mathcal{R} = \emptyset$\\ $\V \gets$ \emph{Derivatives} \\ \emph{estimation}$(\Y)$} ;

    \node [rectangle, align=center, rounded corners, inner
        sep=0pt, minimum height=1.5cm, minimum width=2.5cm, below of=velo, node
        distance=1.85cm,draw=darkblue]
        (skeleton) {$C \gets$ \emph{Skeleton} \\ \emph{Generation}\\$(\V, \textcolor{red}{\delta_{\max}})$};

    \node[rectangle, align=center, rounded corners, ultra thick,
        inner sep=0pt, minimum height=1.5cm, minimum width=2.5cm, below
        of=skeleton, node distance=2cm, draw=darkblue]
        (kinetic) {$C \gets$ \emph{Kinetic}\\ \emph{Inference}\\$(C, \Y, \V)$} ;

    \node[diamond, align=center, rounded corners, inner sep=0pt,
        minimum height=1.5cm, minimum width=2.5cm, right of=kinetic, node
        distance=3.75cm,draw=darkblue]
        (select) {$r^* \gets$ \emph{Selection}\\ \vspace{.05cm}$(C,
        \textcolor{red}{\alpha})$} ;

    \node [rectangle, align=center, rounded corners, inner
        sep=0pt, minimum height=1.2cm, minimum width=3.3cm, below of=select,
        node distance=3cm,draw=darkblue]
        (update) {$\mathcal{R} \gets \mathcal{R} \cup \{r^*\}$ \\ \vspace{.05cm}$\V \gets$ \emph{Update}$(\V, r^*)$};

    \node [rectangle, align=center, rounded corners, ultra thick,
        inner sep=0pt, minimum height=1.5cm, minimum width=2.5cm, right
        of=select, node distance=3.9cm, draw=violet]
    (polish) {$\mathcal{R} \gets$ \emph{Global}\\ \emph{Optimization}\\ $(\V^{\text{init}}, \mathcal{R})$};

\node [rectangle, align=center, rounded corners, ultra
        thick, inner sep=0pt, minimum height=1.2cm, minimum width=2.5cm, below
        of=polish, node distance=3cm, draw=red!60!black]
        (output) {Output \\ CRN $\mathcal{R}$};

\draw [post] (input) -- (velo);
\draw [post] (velo) -- (skeleton);
\draw [post] (skeleton) -- (kinetic);
\draw [post] (kinetic) -- (select);
\draw (select) edge [post, text width=1.5cm, below, pos=-.1] node {yes} (update);

\draw (select) edge [post, text width=2.2cm, above, pos=1.1] node {no} (polish);

\draw[black, thick, ->] (update.west) -- ++(-130pt,0pt) |- (skeleton.west);

\draw [post] (polish) -- (output);
\end{tikzpicture}};
}

    \node[] at (2.5,-0.5) {
    \begin{tikzpicture}[every node/.style={scale=.4}]

                    \node[rectangle, fill=white, draw=black] (root) {\huge{$\emptyset$}};


    \node[below left=1cm and 3cm of root] (r11) {\LARGE{$A\reac{1}C$}};
    
    \draw (root) edge [treeline, text width=2.6cm, above, pos=.5, font=\LARGE]
        node {} (r11.north);

    \node[right of=r11, node distance=4.75cm, font=\LARGE] (r12) {$A\reac{1}C+D$};
    
    \draw (root) edge [treeline, text width=2.3cm, left, pos=.5, font=\LARGE]
        node {} (r12.north);

    \node[right of=r12, node distance=5.5cm, font=\LARGE] (r13) {$A\reac{1}B+C$};
    
    \draw (root) edge [treeline, text width=2.8cm, above, pos=.5, font=\LARGE]
        node {} (r13.north);

    \node[right of=r13, node distance=16cm, font=\LARGE] (r14) {$D\reac{1}E$};
    
    \draw (root) edge [treeline,red, text width=2.8cm, above, pos=.5, font=\LARGE]
        node {} (r14.north);

    \node[below of=r11, font=\LARGE, node distance=2cm] (r21) {$D\reac{1}E$};
    
    \draw[treeline] (r11.south) to (r21.north);

    \node[below of=r12, node distance=2cm, font=\LARGE] (r23) {$D\reac{1}E$};
    
    \draw[treeline] (r12.south) to (r23.north);

    \node[below of=r13, node distance=2cm, font=\LARGE] (r26) {$D\reac{1}E$};
    \draw[treeline] (r13.south) to (r26.north);

    \node[right of=r26, node distance=9cm, font=\LARGE] (r29) {$A\reac{1}B$};
    \draw[treeline] (r14.south) to (r29.north);

    \node[right of=r29, node distance=11.5cm, font=\LARGE] (r28) {$C\reac{1}D$};
    \draw[treeline, red] (r14.south) to (r28.north);




    \node[below of=r21, font=\LARGE, node distance=2cm] (r31) {$C\reac{1}D$};
    \draw[treeline] (r21.south) to (r31.north);

    \node[below of=r23, node distance=2cm, font=\LARGE] (r33) {$C\reac{0.95}D$};
    
    \draw[treeline] (r23.south) to (r33.north);

    \node[right of=r33, node distance=3.5cm, font=\LARGE] (r35) {$C\reac{1}D$};
    
    \draw[treeline] (r26.south) to (r35.north);

    \node[right of=r35, node distance=3cm, font=\LARGE] (r39) {$B\reac{1}D$};
    
    \draw[treeline] (r26.south) to (r39.north);

    \node[below of=r29, node distance=2cm, font=\LARGE] (r38) {$B\reac{1}C+D$};
    
    \draw[treeline] (r29.south) to (r38.north);

    \node[right of=r38, node distance=10.5cm, font=\LARGE] (r37) {$B\reac{1}C$};

    \draw[treeline,red] (r28.south) to (r37.north);

    \node[right of=r37, node distance=8cm, font=\LARGE] (r36) {$A\reac{1}B$};

    \draw[treeline] (r28.south) to (r36.north);

    \node[below of=r33, font=\Large, node distance=2cm] (r43) {$B\reac{0.70}C$};

    \draw[treeline] (r33.south) to (r43.north);

    \node[below of=r31, node distance=2cm, font=\Large] (r41) {$B\reac{1}C$};
    
    \draw[treeline] (r31.south) to (r41.north);

    \node[below of=r35, node distance=2cm, font=\Large] (r45) {$B\reac{1}C$};
    
    \draw[treeline] (r35.south) to (r45.north);

    \node[below of=r39, node distance=2cm, font=\Large] (r49) {$C\reac{0.81}D+E$};
    
    \draw[treeline] (r39.south) to (r49.north);

    \node[right of=r49, node distance=5cm, font=\Large] (r47) {$C\reac{1}D+E$};
    
    \draw[treeline] (r38.south) to (r47.north);

    \node[right of=r47, node distance=6cm, font=\Large] (r46) {$C\reac{1}D$};
    
    \draw[treeline] (r38.south) to (r46.north);

    \node[below of=r36, node distance=2cm, font=\Large] (r400) {$B\reac{1}C+E$};
    
    \draw[treeline] (r36.south) to (r400.north);

    \node[below of=r37, node distance=2cm, font=\Large] (r401) {$A\reac{1}B+E$};
    
    \draw[treeline] (r37.south) to (r401.north);

    \node[left of=r401, node distance=4cm, font=\Large] (r402) {$A\reac{1}\emptyset$};
    
    \draw[treeline] (r37.south) to (r402.north);

    \node[right of=r401, node distance=4cm, font=\Large] (r403) {$A\reac{1}B$};
    
    \draw[treeline, red] (r37.south) to (r403.north);

\end{tikzpicture}
};
\node at (-6, 8) {\textbf{a}};
\node at (-6, 1.5) {\textbf{c}};
\node at (2.2, 8) {\textbf{b}};

\end{tikzpicture}
 \caption{\textbf{Overview of Reactmine execution on one simulation trace of the chain CRN $A\overset{1}\rightarrow B$, $B\overset{1}\rightarrow C$, $C\overset{1}\rightarrow D$, $D \overset{1}\rightarrow E$.}
\textbf{a} Reactmine flowchart along one path of the search tree with final global re-optimization.   
\textbf{b} Input simulation trace with marking of the reaction supports of the best ranked CRN.
\textbf{c} Part of the search tree corresponding to the 10 best ranked CRNs after global re-optimization (the best ranked CRN branch is in red).}

\label{fig:chain}
\end{figure*}

In this paper which extends~\citep{MGSF19cmsb}, we present Reactmine, a bounded-depth tree search algorithm to infer CRNs from time series data,
without any low correlation assumption.
Sparsity is enforced by inferring reactions with their kinetics in a sequential fashion,
with the depth of the search tree bounding the number of inferred reactions.
At each node, the reaction candidates are ranked according to the variance of the kinetics inferred on their transition support,
and the best candidates are used as choice points at that node.
At each successor node, one selected reaction is added and its effect is subtracted from the trace.
Each leaf in the search tree represents a CRN candidate.
In a final pass, the kinetic parameters of the leaf CRNs are globally re-optimized on the whole trace transitions,
and the CRN candidates are ranked according to the quadratic loss between the predicted and experimentally-observed temporal variations.
As an example, \Cref{fig:chain}\textbf{c} shows the recovery by Reactmine of the chain CRN $A \reac{} B \reac{} C \reac{} D \reac{} E$
with mass action law kinetics and rate constants equal to one, from a single simulation trace of $A, B, C, D$ and $E$
with $A$ initially present.

On a benchmark of synthetic data obtained by simulation from a hidden CRN with standard initial conditions,
we show the capability of Reactmine to recover either the hidden CRN, or a variant CRN capable of reproducing the simulation data
in the same range of initial conditions, whereas SINDy fails to infer sparse ODE systems
and even to reproduce the time series data on different initial conditions.
In these examples, we analyze the sensitivity of the results to the number of time points and to the four hyperparameters of our algorithm:
$\gamma$, the maximum number of reactions inferred (i.e.\ maximum depth of the search tree),
$\beta$, the maximum number of reaction candidates (i.e.\ maximum branching factor of the search tree),
$\delta_{\max}$, the maximum difference of concentration change between species taking part in a reaction candidate,
and $\alpha$, the coefficient of variation acceptance threshold about the inferred kinetics on the supporting transitions.

Then, we apply Reactmine on two sets of real experimental data: one from protein
fluorescence videomicroscopy of cell cycle and circadian clock markers in mammalian fibroblasts,
and one from biomedical measurements of systemic circadian biomarkers
possibly acting on clock gene expression in peripheral organs.
We show that Reactmine succeeds in inferring meaningful interactions,
interestingly in accordance with the main conclusions drawn from previous analyses of these datasets though ODE models
built, respectively, using a temporal logic approach in ~\cite{TFS16biosystems}, and a different model learning approach in ~\cite{MDL21bioinformatics}.

The rest of the article is organized as follows.
In the Methods section, we present our CRN inference algorithm, its theoretical complexity and comparison to related work.
In the Results section, we first evaluate its performance on synthetic data obtained by simulation of some hidden CRNs,
and perform the above-mentioned sensitivity analyses.
Then we show the results obtained with simulation data from 
the MAPK signaling CRN studied in~\cite{QNK07ploscb}.
In all those instances, our results are shown to compare favorably to SINDy.
Then, we present our results on the two real-world biological datasets of this study, 
and compare them to the previous models developed from those data.
Finally, we conclude on the merits of this approach and its current limitations.

\section{System and methods}

In this article, bold lower (resp.\ upper) case letters denote vectors (resp.\ matrices).
Unless stated otherwise, sets are represented with capital letters.
For a matrix ${\bf{M}}$,~$\M_{i, \bullet}$ (resp.\ $\M_{\bullet, j}$)
stands for its $i^{\text{th}}$ row (resp.\ $j^{\text{th}}$ column).

\subsection{Temporal data}

We are given the temporal evolution of the concentrations of $m$ biological species
at $n$ discrete time points $\{t_i\}_{1\leqslant i\leqslant n}$, stored in a matrix
$\Y\in\R^{n\times m}$
coming from either a single experiment or multiple traces.
We consider the matrix
$\V\in\R^{n\times m}$ of concentration change velocities obtained by
finite difference approximation of the derivatives:
$v_{i,j}=\frac{y_{i+1,j}-y_{i,j}}{t_{i+1}-t_i}$ and  $v_{i,j}=\frac{y_{i-1,j}-y_{i,j}}{{t_{i-1}-t_i}}$ for the last point of a trace,
and more precisely the matrix of normalized concentration change velocities $\tilde{\V}$ defined by $\displaystyle{\tilde{v}_{i,j} = \frac{v_{i,j}}{\max_{i\in \{1,\dots,n\}}~\lvert v_{i, j}\rvert}}$.

\subsection{Chemical Reaction Networks (CRN)}

A chemical reaction is formally defined as a triple $(R, P, f)$,
where $R$ (resp.\ $P$) is a multiset of reactant (resp.\ product) species and
$f:\R^n_+ \to \R_+$ is a rate function over molecular concentrations specifying the reaction kinetics.
A reaction catalyst is a species in $R\cap P$.
A chemical reaction network (CRN) is a finite set of reactions. 

For the sake of simplicity in the presentation, we restrict ourselves to reactions with 0/1 stoichiometry only,
and do not describe the handling of other reaction patterns such as autocatalysis.
The stoichiometric change vector of a reaction is thus a vector $\bs \in \{-1, 0, 1\}^m$  defined by 
$s_j =1 \text{ if } j \in P \setminus R, -1 \text{ if } j \in R \setminus P, 0$ otherwise.
A reaction with mass action law kinetics with rate constant $k$ is written $R \overset{k}\rightarrow P$.
Michaelis-Menten and Hill kinetics are also considered in the inference algorithm.

\subsection{Reactmine CRN inference algorithm}

Reactmine is a bounded-depth tree search algorithm which, at each node of the search tree, 
\begin{enumerate}
\item
infers reaction candidates composed of the reactants and products with the highest change at some observed transitions called their support,
\item
ranks the reaction candidates according to the variance of the ratio between the observed and the inferred kinetics on their support,
\item
and selects the $\beta$ best reaction candidates by adding them as successor nodes and substracting their effect on the velocity matrix.
\end{enumerate}
At the end, a global re-optimization of the kinetic parameters of the inferred CRNs at the leaves of the search tree is performed on the whole trace, 
in order to select the inferred CRN that minimizes the quadratic loss between the inferred and observed concentration change velocities on the data.
Reactmine uses four hyperparameters:
\begin{itemize}
\item $\gamma$, the maximum depth of the search tree, i.e.~the maximum number of inferred reactions in a CRN candidate,
\item $\beta$, the maximum number of reaction candidates considered at a node,
\item $\delta_{\text{max}}$,  the maximum concentration change ratio allowed for inferring the reactants and products of one reaction at a given transition,
\item  $\alpha$, the maximum kinetics variance allowed on the support of a reaction. 
\end{itemize}
The different phases of Reactmine are shown in Fig.~\ref{fig:chain} and detailed below.

\subsubsection{Inference of reaction skeletons}

Let $\overline j={\argmax_{j\in\{1,\dots,m\}}}~\lvert \tilde{v}_{i,j}\rvert$ be the index of the species having the highest concentration change at time $t_i$.
That species $\overline j$ is assumed to undergo the primary change that should be explained
by one preponderant reaction at that time point.

We consider the set $r(i)=\{(R_\delta(i),P_\delta(i))~|~\delta\in\Delta\}$
 of reaction skeletons involving reactants and products that have concentration changes similar to species $\overline j$ at time $t_i$, defined by

\begin{equation}
    R_\delta(i) = \{j \in \{1, \dots, m\}~|~ v_{i, j} < 0,~\left\lvert{\frac{v_{i, {\overline j}}}{v_{i,j}}}\right\rvert \leq {\delta}\}
\label{eq:deltamax} 
    \end{equation}
$$P_\delta(i) = \{j \in \{1, \dots, m\}~|~ v_{i, j} > 0,~\left\lvert{\frac{v_{i, {\overline j}}}{v_{i,j}}}\right\rvert \leq {\delta}\}$$
for a finite set $\Delta$ of $\delta$ ratios comprised between $1$ and  $\delta_{\text{max}}$ hyperparameter value (typically 3).
Note that species ${\overline j}$ belongs to the reaction skeleton.
In the particular configuration $r_\delta(i) = (\emptyset, \{{\overline j}\})$,
a synthesis reaction $\emptyset \rightarrow {\overline j}$ is obtained. Likewise
$r_\delta(i) = (\{{\overline j}\}, \emptyset)$
leads to a degradation reaction ${\overline j} \rightarrow \emptyset$.
This setting allows us to infer several reaction skeleton candidates,
e.g.\ for $\delta' > \delta$ one could have $r_\delta(i) = (\{A\},\{{\overline j}\})$
and $r_{\delta'}(i) = (\{A, B\}, \{{\overline j}\})$.
A value of $\delta$ significantly above 1 accounts for the fact that
$v_{i, {\overline j}}$ may not be completely explained by one reaction only, but by some other reactions which will be discovered later on.

The support $\T(r)$ of a reaction skeleton $r=(R,P)$ is then defined 
as the set of time
points indices where it is infered: 
    \begin{equation}
        \T(r) = \left\{i \in \{1, \dots, n\}~|~ \exists \delta \in \Delta, ~ r_{\delta}(i) = (R, P)\right\}
    \end{equation}

\subsubsection{Statistical inference of reaction kinetics}\label{subsec:kinetic}

The next step consists in assigning a rate function to the reaction skeleton, to completely define one reaction. We here describe the process for mass action law reactions. Michaelis Menten and Hill kinetics reactions are covered in Supplementary Section~\ref{sec:suppmm}.

$(R, P, f)$ follows the law of mass action with
parameter $k$ if
$\forall j \in R \cup P,~ \forall i \in \{1, \dots, n\}$
\begin{equation}
v_{i, j} = s_j f(\Y_{i, \bullet}) =  s_j k \prod_{u \in R} y_{i, u}
\label{eq:MAL}
\end{equation}

\noindent where we recall that $s_j$ is the stoichiometry of species $j$ in the reaction.
Using the finite differences estimate $\V$ as well as the
support set $\T(r)$ for the current reaction candidate $r=(R,P)$,
one can provide an estimator of $k~\forall j \in R \cup P$:

\begin{equation}
    \hat{k}_j = \frac{s_j}{\#\T(r)} \sum_{i \in \T(r)}
    \frac{v_{i, j}}{\prod_{u \in R} y_{i, u}}
\label{eq:hatk}
\end{equation}

\noindent This estimator is designed to realize an
equality in mean across the support between observed kinetics and inferred kinetics.
The former is represented by the numerator, the latter by the denominator times $\hat{k}_j$.

\label{subsec:selection}

In order to compare reaction candidates between them, an interesting criterion
to look at is certainly the statistical quality of the inferred kinetics on the support of the inferred reaction skeleton.
The variance $\sigma_j$ of the mass action law coefficient estimate over the support of the reaction can be estimated itself
for each species involved in the reaction:
It is worth noticing however that there is a relationship between mean
and variance when estimating the kinetics of different reactions:
a slow reaction will tend to produce a low variance, compared to a faster
reaction.
We thus consider
the coefficient of variation (CV),
$\rho_j = \frac{\sigma_j}{\lvert\hat{k}_j\rvert}$
measured for each reactant or product of the reaction,
and introduce more precisely the species index that minimizes it:
$j^* = \underset{j \in R\cup P}{\argmin}~\rho_j$
on which we rely to estimate $k$. The complete reaction is therefore $r = (R, P, \hat{f})$ with
$\displaystyle{\hat{f} :\y \mapsto \hat{k} \prod_{u\in R}y_u}$ and $\hat{k} = \hat{k}_{j^*}$.
This process is performed for all reaction skeleton candidates.

A reaction candidate $r$ is \emph{accepted} if it satisfies $\rho(r) < \alpha$.
A typically acceptable value for $\alpha$ is below 1,
indicating that the variance of the estimator does not overcome the mean.
In the event where the best reaction $r^*$ fails to satisfy that condition,
 the addition of a catalyst to the reaction candidate is tried.
To that end, Equation~(\ref{eq:MAL}) is modified:

\begin{equation}
    v_{i,j} = s_j k \prod_{u \in R \cup \{c\}} y_{i,u}
\label{eq:MALcata}
\end{equation}
$\forall j \in R\cup P$, and $c$ can be any species. The
optimal catalyst $c^*$ for a particular reaction is the species providing the
lowest CV, in which case $R \gets c^*$ and $P \gets c^*$.
A catalyzed reaction is accepted if
its associated CV is below $\alpha$.

The reaction are thus ranked according to their CV, the lowest CV corresponding
to the best reaction.
The $\beta$ best accepted reactions
are returned, representing the maximum
number of inferred candidates.

\subsubsection{Update of the matrix of concentration change velocities}\label{algo:update}

Once a candidate reaction $(R,P, f)$ is accepted and selected as successor node in the search tree,
  its effect on the velocity matrix is substracted at that node as follows:
  
\begin{equation}
    \V \gets \V -
    \begin{pmatrix} f(\Y_{1, \bullet}) \\
    \vdots
    \\ f(\Y_{n, \bullet})
\end{pmatrix} \bs^T
\label{eq:velupdate}
\end{equation}

It is worth remarking that that substraction operation can also be used in our approach
in order to take into account prior knowledge consisting of reactions already known to act on the observed species,
simply by applying it on the input traces as a preprocessing step.

\subsubsection{Theoretical complexity}

\noindent \textbf{Proposition.}
The computational time complexity to infer one reaction $(R, P, f)$ is $\mathcal{O}(nmJ)$
where $n$ is the number of time points, $m$ the number of species, and $J = |R\cup P|$.
The global time complexity of Reactmine is $\mathcal{O}(\beta^\gamma nmJ)$.

\noindent \textbf{Proof.}
Inferring the reaction kinetics constant involves the computation of a mean
for each species present in the reaction (Equation~\ref{eq:hatk}), which is
$\mathcal{O}(nJ)$. In the worst-case, a lookup for a catalyst species is
necessary, at a cost of $\mathcal{O}(nJm)$. The update of velocities performed
in Equation~(\ref{eq:velupdate}) is
$\mathcal{O}(nJ)$. Generating reaction skeletons requires
the computation of the species displaying highest absolute variations for each time
point, which is $\mathcal{O}(nm)$ 
After that, the
sets $R_\delta(i)$ and $P_{\delta}(i)$ are obtained with a bounded number of $\delta$ values.
The time complexity for the inference of one reaction is therefore $\mathcal{O}(nmJ)$.

Since the depth of the search tree is bounded by $\gamma$ and each node has at most $\beta$ children,
the time complexity of Reactmine is thus $\mathcal{O}(\beta^\gamma nmJ)$.

\subsubsection{Final global re-optimization of kinetic parameters}\label{subsec:polish}

During search, the kinetics of the reactions are estimated on their limited support.
At the end of the search, the kinetics parameters of each CRN candidate
 can be corrected  by global re-optimization on the whole input trace for taking into account the effect of all reactions concurrently.
Given an inferred CRN $\mathcal{R} = \{(R_q, P_q, f_q)\}_{1 \leqslant q\leqslant p}$ composed of $p$ reactions
and data matrix $\Y$, one can construct a matrix
$\bF(\Y, \bk) :=  (f_q(\Y_{i, \bullet}, \bk))_{\substack{1\leqslant i \leqslant n\\ 1\leqslant q \leqslant p}} \in \R^{n\times p}$
with $n$ the number of time points.
The $q^{\text{th}}$ column of $\bF(\Y, \bk)$ is a vector describing
the rate of the reaction $(R_q, P_q, f_q)$ at each time point, with $\textbf{k}$ being the vector of reaction kinetic parameters.
Combined with the stoichiometry matrix of the CRN $\bS \in \R^{p\times m}$,
we can formulate the optimization problem:

\begin{equation}
    \bk = \underset{\bk \in \R^p_+}{\argmin}~\norm{\V -
    \bF(\Y, \bk)\bS}_F^2
\label{eq:polish}
\end{equation}

\noindent where $\norm{\cdot}_F$ is the Frobenius norm. E.g.~for a mass action law CRN,

\begin{equation}
    \bF(\Y, \bk) = \begin{bmatrix} | &  & | &  & | \\ \underset{j \in R_1}{\prod} y_{i, j} & \dots & \underset{j \in R_q}{\prod} y_{i,j} & \dots &
        \underset{j \in R_p}{\prod} y_{i,j}\\ | & & | & & |\end{bmatrix} \text{diag}(\bk)
\label{eq:convex}
\end{equation}

The optimization starts with an initial guess set as $\textbf{k}$ $= (k_1,\dots,k_p)^T$.
It is worth noticing that the least squares term compares the inferred and
observed derivatives, rather than the data measurements $\Y$ and a numerical
integration of the inferred CRN, which allows avoiding the resolution of a
non-convex optimization problem in the case of mass action law kinetics. Indeed, Equation~(\ref{eq:convex})
shows that the inferred velocities are written as a weighted linear combination
of reaction effects, which makes the minimization problem convex.
Furthermore, in order to take into account
the fact that concentrations might span a wide range from one species to another,
we normalize the $j^{\text{th}}$ column of the matrix $(\V -
\bF(\Y, \bk)\bS)$ by $\underset{i \in 1\le i \le n}{\max} \V_{i,j}$,
for all $j$, inducing equal importance for each species in the cost function.
This final step of global optimization is performed for all CRN candidates at the leaves
of the search tree.
Reactmine returns the CRN which minimizes the loss function defined in Equation~(\ref{eq:polish}).

More globally, we use the same loss function to find values for the hyperparameters of Reactmine by gridsearch,
as described in the forthcoming sensitivity analysis section.

\subsection{Related work}\label{subsec:othersSINDy}

Some methods
originally designed to discover the dynamics of physical systems can be applied to
our problem of inferring a CRN from time series data.
Most notably, the SINDy system \citep{BPK16pnas}, starting from temporal measurements,
aims at providing a reconstruction of the velocities in the following way:

\begin{equation}
    \V = \bTheta(\Y)\bXi
\label{eq:SINDymod}
\end{equation}

$\bTheta(\Y) \in \R^{n\times p}$ is a library of $p$ functions
constructed from the input variables $\Y$ including, for instance, first to $m$-order
polynomial interactions, e.g.\ $\Y_{\bullet,j}\odot \Y_{\bullet,j'}$ the $\sin$ and $\cos$ functions,
e.g.\ $\sin(\Y_{\bullet,j})$,
or even more sophisticated user-defined functions.
The dynamics of each variable is then captured by a weighted combination of library members,
the weights being encompassed in $\bXi$. Because it is thought that the expression of the
dynamics should be sparse within the library $\bTheta(\Y)$, SINDy
proposes to obtain $\bXi$ using sparse regression.
\begin{equation}
\bXi = \underset{\bXi \in \R^{p\times m}}{\argmin}~\norm{\V -
    \bTheta(\Y)\bXi}_F^2  + \lambda \norm{\bXi}_1
\label{eq:SINDyopt}
\end{equation}

$\lambda$ is an hyperparameter governing the tradeoff between goodness-of-fit
and sparsity.
For fair comparison within our CRN setup with mass action law and
stoichiometry at most 1, $\bTheta(\Y)$ is here restricted to polynomials up to
the second order, and a bias term. Hence $p= \left(1+\frac{m(m+1)}{2}\right)$.

\begin{equation}
    \bTheta(\Y) :=
    \begin{bmatrix}
        | & | &  & | & | & & | \\
        1 & \Y_{\bullet, 1} & \dots & \Y_{\bullet, m} & \Y_{\bullet,1}\Y_{\bullet,2} & \dots & \Y_{\bullet, m-1}\Y_{\bullet, m}\\
        | & | &  & | & | & & | \end{bmatrix}
\label{eq:thetaSINDy}
\end{equation}

Associated to a positive weight, the bias term corresponds to a synthesis
reaction.
First order interactions translate to reactions such as $A\overset{k}{\rightarrow}
B +C$ for $A \neq \emptyset$, with the special case $A \in \{B, C\}$ corresponding to
a catalyzed synthesis. Second order interactions encompass reactions of the form
$A+B\overset{k}{\rightarrow} C+D$ for $\{A, B\} \neq \emptyset$. Again, the case
$A \veebar B \in \{C, D\}$ corresponds to a catalyzed reaction, with $\veebar$
referring to the exclusive OR\@.
In our experiments, we use the \texttt{pySINDy} package with STLSQ optimizer, as
the latter yielded the best results.

\section{Results}

\begin{table*}[ht!]
    \centering
    \resizebox{\textwidth}{!}{
\begin{tabular}{llc}
                \toprule
                Hidden CRN & CRN inferred by Reactmine & ODE system inferred by SINDy \\
                \midrule
                 \makecell[l]{Chain CRN\\  \\$\textbf{A}\reac{1} B$\\$B\reac{1} C$\\
                $C\reac{1}D$\\$D\reac{1}E$}
                           &
                           \makecell[l]{$\gt{D\reac{1.00} E}$\\[8pt]$\gt{C\reac{1.00} D}$\\[8pt]$\gt{B\reac{1.00} C}$\\[8pt]$\gt{A\reac{1.00}
                           B}$\\[5pt]}
                           &
                           \makecell[l]{$\left\{\begin{array}{l}
                            \begin{aligned}
                                \dot{A} &= \gt{-1.00 A}\\[-7pt]
                                \dot{B} &= \gt{1.00 A -1.00 B}\\[-7pt]
                                \dot{C} &= \gt{1.03B-1.03C}~ \ft{+0.01D-0.06AB}\\[-7pt]
                                \dot{D} &= \ft{0.33B-0.64DE}\\[-7pt]
                                \dot{E} &= \gt{1.00D}
                            \end{aligned}\end{array}\right.$} \\ 
                \midrule
               \makecell[l]{MAPK CRN\\  \\$\textbf{A}\reac{0.0045}Ap$\\ \\$Ap+\textbf{B}\reac{1000} ApB$\\ \\$ApB\reac{150} Ap+\textbf{B}$\\ \\$ApB\reac{150}
Ap+Bp$\\ \\$Ap+Bp\reac{1000} ApBp$\\ \\$ApBp\reac{150}Ap+Bp$\\ \\$ApBp\reac{150}Ap+Bpp$}
            &\makecell[l]{$\gt{A\reac{0.0045}Ap}$\\[8pt]
            $\ft{Bp+Ap\reac{499.97} Bpp+Ap}$\\[8pt]
            $\at{B+Ap\reac{500.01} Ap}$\\[8pt]
            $\gt{ApB\reac{150.04}Ap+Bp}$\\[8pt]
            $\gt{Ap+B\reac{501.19} ApB}$\\[8pt]
            $\gt{ApB\reac{150.37}Ap+B}$\\[8pt]
            $\gt{Ap+Bp\reac{517.78}ApBp}$\\[8pt]
            $\gt{ApBp\reac{155.34}Ap+Bp}$\\[8pt]
            $\at{Ap+B\reac{500.19}ApB+B}$\\[8pt]
            }                                 
                          &
            \makecell[c]{\makecell[l]{No sparse ODE system with a good loss function is inferred  for any value of $\lambda$}\\ \\
            \includegraphics[width=.6\linewidth]{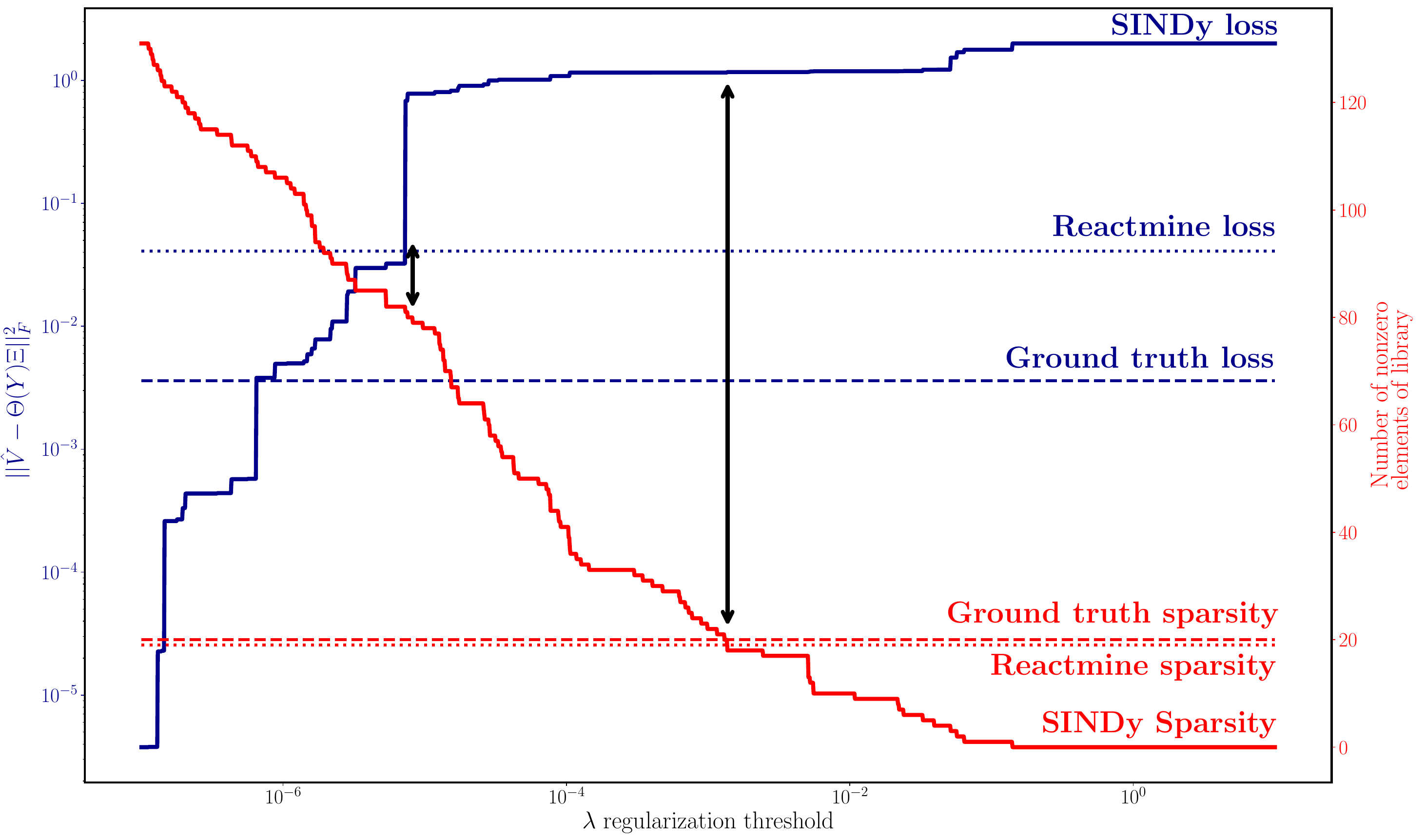}}
            \\
        
            \bottomrule
\end{tabular}
    }
\caption{\textbf{Results obtained by Reactmine and SINDy on the Chain and one level MAPK CRNs} using a single simulation trace from one initial state
containing the molecular species indicated in bold in the first column.
The reactions learned by Reactmine are indicated in green if they belong to the hidden CRN structure,
in yellow if they lead to equivalent terms of the associated ODEs, and in red otherwise. 
For the ODE systems inferred by SINDy on the Chain example, the terms are colored in green if they correspond to the kinetics of some hidden reactions, regardless of the precise kinetic constant value as long as the sign is exact,
and in red otherwise. 
For the MAPK CRN, SINDy fails to learn the hidden dynamics (see Table~\ref{tab:results} where other examples are given).
The lower right panel reports the quadratic training loss (blue) and the number of nonzero
terms in the library (red) found by SINDy, as a function of $\lambda$.
The dashed red line represents the actual
number of nonzero terms in the ground truth CRN. The dotted blue line stands for the quadratic loss
value found by Reactmine. The left arrow shows that the ODE model inferred by SINDy
matching the Reactmine loss contains too many terms, around 90,
while the right arrow demonstrates that the sparse ODE systems inferred by SINDy are not able to fit the input trace.
}
\label{tab:resultsmain}
\end{table*}

\subsection{Evaluation on simulation data from hidden CRNs}\label{subsec:evalhidden}

We first evaluate our algorithm on simulation data obtained from hidden CRNs,
with which the inferred CRNs can be compared.
For Reactmine, we report the reactions in the order of their inference along the search tree branch that gives the best CRN.
For SINDy, we report the inferred ODE systems. 
In these experiments, numerical integration is performed using the Python package
\texttt{scipy.integrate.odeint} with the default integrator \texttt{lsoda}.
Simulations run for a time horizon $T=10$ and a time step $\Delta t = 0.1$.

Reactmine uses four hyperparameters which are optimized by grid search
to minimize the quadratic loss criterion of Equation~(\ref{eq:polish}).
For all the CRNs considered through the section, we search
$\alpha \in [0.005, 0.5], \beta \in \llbracket 4,8 \rrbracket$ and
$\gamma \in \llbracket 3,6 \rrbracket$, except for the MAPK example where we search for
$\alpha \in [0.0025, 0.2], \beta \in \{6,8,10\}$ and
$\gamma \in \llbracket 7,10 \rrbracket$. We do not search for an optimal
$\delta_{\max}$ and set its value to $3$ as a later shown sensibility analysis will
demonstrate that this hyperparameter is quite unsensible.  
SINDy uses one hyperparameter, the sparsity-enforcing penalty coefficient $\lambda$. 
It turns out that in all the examples below, there is no value of $\lambda$ leading to
both a good fit and a sparse model. 
In particular, there is no value of $\lambda$ for which the hidden dynamics are recovered, as illustrated by Fig.~\ref{fig:sindy}.
For the sake of comparison to SINDy, we thus report the ODE system obtained using the (greatest) value of $\lambda$ that gives the same quadratic loss as Reactmine.

On the chain CRN example, Table~\ref{tab:resultsmain} shows that
Reactmine succeeds in recovering the hidden CRN by inferring $D\reac{1}E$ first,
with the end of the trace as support, (Fig.~\ref{fig:chain}\textbf{c}).
Then the other reactions are learned in backward order, after successive derivative matrix updates, 
for subtracting the effects of each learned reaction.
More details are given in Table~\ref{tab:results} where we see that, in this example, the reactions are immediately inferred with the right kinetics
and not changed by global re-optimization.
Table~\ref{tab:computgrid} gives the hyperparameter settings used for the results reported in this section,
the number of CRN candidates computed using the best hyperparameter setting found (here 128 chain CRN candidates) 
the learning time (here 0.31 seconds) and the hyperparameter grid search computation time (here 50 minutes).

In this example, SINDy  correctly infers the (ODE terms of the) hidden reaction
$A\reac{1}B$, then a part of reaction $B\reac{1}C$ is present: $\dot{B}$ includes
the term $-1.00B$, but only $0.18B$ can be found in $\dot{C}$ among several other terms that do not correspond to reactions.
Likewise, the production of
$E$ by the reaction $D\reac{1}E$ is correctly inferred, but the ODE associated with $D$ is $-0.37DE$ instead of $-1.00D$.
Moreover, the learned ODEs are not able to generalize the dynamics of species $D$ and $E$ on traces
that were not used during training (Fig.~\ref{fig:sindy_nogen_chain}).
It is worth noting in this respect that the chain CRN is composed of 8 ODE terms,
whereas the SINDy library comprises $5 \times (1+\frac{5\times 6}{2})=80$ terms in this example.

The second example concerns the MAPK signaling network, a ubiquitous CRN structure that is present in all eukaryote cells and in several copies.
We consider the simplified two-stage (instead of three) CRN model composed of $7$ species and $7$ reactions of~\cite{QNK07ploscb}. 
The input species $A$
goes through a first stage of complexation and phosphorylation to produce the phosphorylated form $Ap$
which plays the role of a kinase on $B$ at the second stage to produce the doubly phosphorylated output species $Bpp$.
Signal amplification is caused by the difference of concentrations by several orders of magnitude between the input and the output, (Fig.~\ref{fig:mapkplot}).
For this example, we set trace parameters $T=100$ and $\Delta t = \frac{1}{3}$.

As shown in Table~\ref{tab:resultsmain}, Reactmine recovers $6$ out of $7$ reactions of the hidden CRN and $3$ other reactions: 
$ApB\reac{499.96}Bpp+Ap$, $B+Ap\reac{500.01} Ap$ and $Ap+B\reac{500.19}ApB+B$.
It is worth remarking that by summing the ODE terms of the latter two reactions,
we get the same effect as the already inferred reaction $Ap+B\reac{500}ApB$,
but with the original rate constant value $1000$ by summing the two copies of the reaction.
On the other hand, a variant of the missing reaction $ApBp\reac{150}Ap+Bpp$
is inferred by Reactmine, namely $Ap+Bp\reac{500}Ap+Bpp$,
where the reactant complex, $ApBp$ replaced by its two components.
Fig.~\ref{fig:mapk} shows that this is a very good approximation of the ground truth dynamics,
on the training trace, as well as on the other simulation traces obtained from different initial conditions.
 
On the other hand, the ODE system inferred by SINDy has no term close to the hidden CRN, 
contains two zero-valued differential functions for yet evolving species,
and fails on sparsity with an average number of $15$ terms per non-zero ODE.
A more detailed analysis presented in the lower right figure from Table~\ref{tab:resultsmain} shows that SINDy is unable to
produce sparse solutions displaying low error for any value of $\lambda$. In particular, the ODE systems associated with
similar loss values as that of the CRN found by Reactmine involves 90 terms. 
It is worth remarking in that Figure that the loss of the ground truth CRN is not zero for the subtle reason that
the loss is computed by an estimation of the derivative matrix by finite differences on a trace,
produced by numerical integration with a more elaborate implicit method, and overfitted by SINDy for low values of $\lambda$.
It is worth noting that the 7 ODEs of the MAPK CRN comprise a total of $20$ terms, whereas
the SINDy library comprises $7 \times (1+\frac{7\times 8}{2}) = 203$ terms.
Likewise, for the
chain CRN,  
this creates challenging sparse regression problems
in the absence of strong independence properties between predictors \citep{ZY06jmlr}.
The better results reported in~\citep{MBP16ieee} may be explained by the recourse to multiple traces with different zeroes in the initial conditions,
similarly to what has been shown in the context of Boolean models in~\citep{CFS17cmsb}.

Table~\ref{tab:results} summarizes the results obtained by Reactmine and SINDy 
on a benchmark of even smaller size CRNs presenting different kinds of difficulties.
The learning traces used in those examples, also including the chain CRN,
are detailed in Figs.~\ref{fig:reacplot}--\ref{fig:loopplot}, together with their estimated velocities.
The loop CRN adds a feedback reaction $E\reac{1}A$ to the chain CRN, leading to the stabilization of all molecular species on some common concentration value.
Reactmine succeeds in recovering the hidden reactions in forward order, directly with the right kinetics.
Here again, SINDy recovers some terms of the two first reactions and of the last reaction, but among many other overfitting terms
which do not generalize to simulation traces obtained from different initial states,(Fig.~\ref{fig:sindy_nogen_loop}).

The reactant-parallel CRN is just composed 
of one catalytic reaction, $A+C\reac{1}B+C$, where the catalyst $C$ is produced by two concurrent reactions $D\reac{2}C$ and $E\reac{1}C$.
Reactmine first infers the preponderant production of $C$ by $D$, then by $E$,
after what the reaction catalyzed by $C$ is correctly inferred.
One can notice that the rate constant first inferred for the reaction $E\reac{}C$ 
has a small value below its final value by two orders of magnitude.
The reason is that right after the inference of $D\reac{}C$, 
reaction $E\reac{1.00}\emptyset$ is inferred prior to $E\reac{0.01}C$.
The global re-optimization of rate constants has for effect in this case to set to $0$ the rate constant of the second reaction,
in favor of the third reaction $E\reac{1}C$ which is thus finally recovered with the right kinetics.
In this example, SINDy infers a wrong ODE system that does not reproduce the learning trace, even by increasing the number of time points (Fig.~\ref{fig:sensitimechainSINDy}).

The product-parallel CRN is the symmetrical case of two concurrent consumptions of the catalyst $C$ with the production of species $D$ and $E$,
on which SINDy similarly fails.
Reactmine first infers the preponderant transformation of $C$ in $E$,
then of $C$ in $D$ and then the correct catalyzed reaction with little correction of the rate constants by the global optimization phase, (Table~\ref{tab:results}).

\begin{figure}
        \includegraphics[width=1\linewidth]{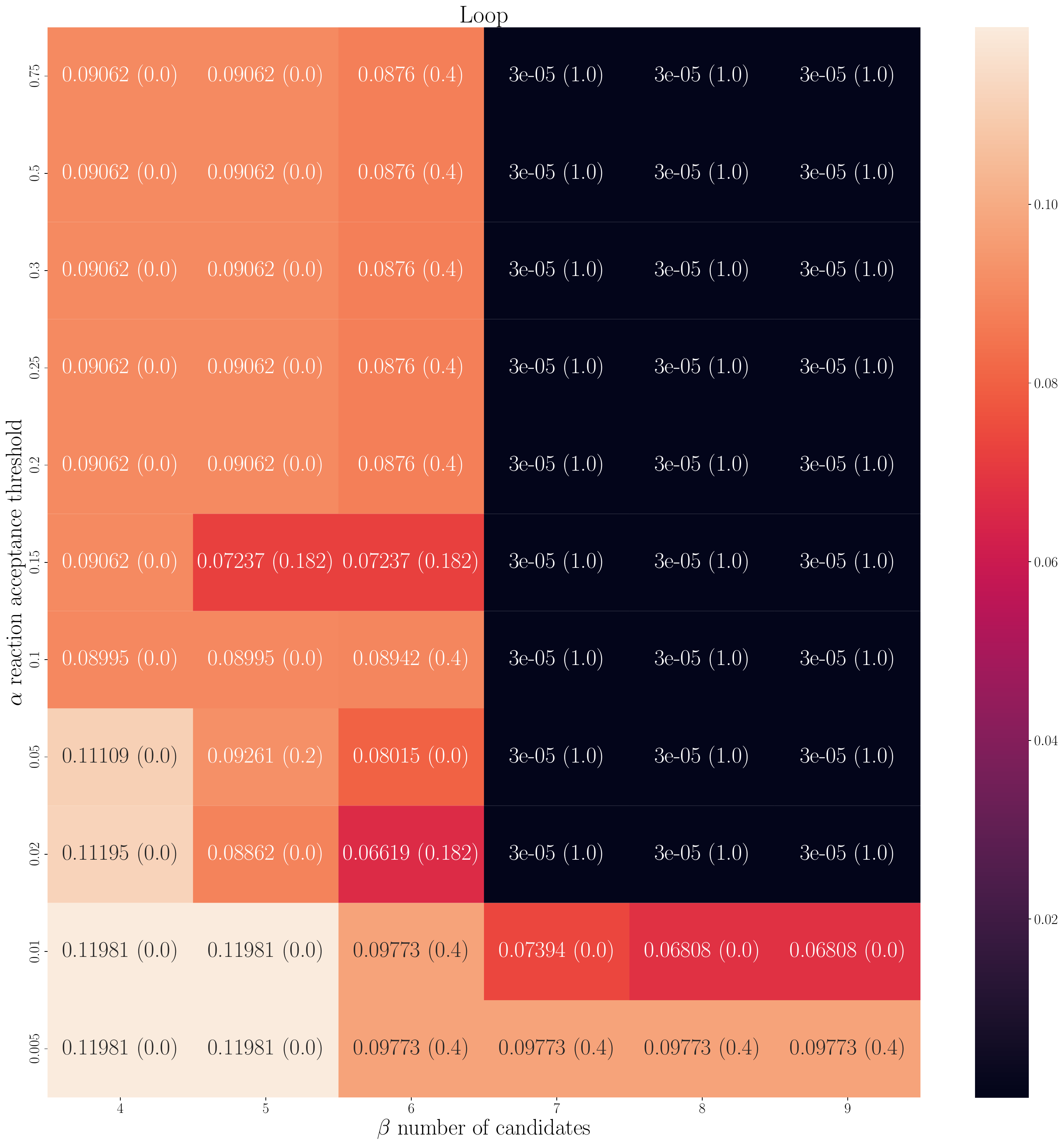}
        \caption{\textbf{Sensitivity of Reactmine to $\alpha$ and $\beta$ hyperparameters for the loop CRN}.
    Quadratic loss $\norm{\V-\bF(\Y,\bk)\textbf{S}}_F^2$ (lower, colored in black, is better) and
     F1-score (in parenthesis, higher is better). The colorbar levels relate to the
     quadratic loss. $\delta_{\text{max}}=3$ and $\gamma=6$.}
\label{fig:sensiloop}
\end{figure}

\subsection{Hyperparameter sensitivity analyses}\label{subsec:sensi}

In this section, we study the impact on the previous resuls of the hyperparameter values of Reactmine
and of the number of time points.
Reactine results are sensitive to both $\alpha$ and $\beta$
(Fig.~\ref{fig:sensiloop} for the loop CRN, Fig.~\ref{fig:supheatmap} for the other toy CRNs).
In particular, there exist a region for which sufficiently many candidates are proposed
and accepted (for the loop CRN, $\beta > 6, \alpha>0.01$).
Next, Fig.~\ref{fig:sensideltagamma} also assesses the sensitivity with respect to
$\gamma$ the maximal CRN size and $\delta_{\text{max}}$ the maximum absolute fold
change between species variations in a reaction.
Only the value of either $\delta_{\text{max}}$ or $\gamma$ is changed, the other hyperparameters are set
to values leading to maximum $F_1$-score. 
$\gamma$
is a sensitive hyperparameter either in terms of quadratic loss value or $F_1$-score.
This is expected as being the maximum CRN size, $\gamma$ has high impact on the flexibility
of the network.
One should remark that the $\gamma$ value associated with a perfect $F_1$-score
is sometimes higher than the size of the hidden CRN. As previously mentioned in
Section~\ref{subsec:evalhidden} for the case of the Reactant-Parallel CRN, this is
due to the inference of reactions whose effect is later set to 0 upon global re-optimization.
For this example, these steps were somewhat needed otherwise the ground truth CRN
would have been recovered with $\gamma=3$.
It is worth noticing that for all examples, the hyperparameter sets yielding the lowest quadratic loss are
associated with highest F1-score for all CRNs, thus providing empirical evidence
concerning the relevance of hyperparameter selection based on quadratic loss minimization.

Now, the sensitivity to the number of time points in the trace is evaluated in
Fig.~\ref{fig:sensitimechain} on the chain CRN. We 
observe an almost monotonic transition to high F1-score / low reconstruction error as the
number of time points increases with optimal performance being reached above 40 time points only.

On the other hand, Fig.~\ref{fig:sensitimechainSINDy} shows that the failure
of SINDy to recover the hidden CRNs persists
independently of the number of time points for both the reactant and product-parallel examples.
For these examples, increasing up to $2\times 10^5$ time points did not lead to perfect inference
as can be seen from the $F_1$-score being different from $1$. The learned models display high precision
but low recall, suggesting sparse but yet incomplete dynamics. However,
the chain and loop CRN could be recovered using $10000$ and $500$ time points, respectively.
This suggests that the problem of sparse regression in that approach
rather comes from the high level of correlations observed in single CRN
traces without the possibility to vary the zeroes in the initial conditions.

\subsection{Evaluation on videomicroscopy data}

Next, we apply Reactmine to time lapse videomicroscopy data 
in NIH-3T3 embryonic mouse fibroblasts \citep{FKT14pnas}, 
used to develop a
coupled model of the cell cycle and the circadian clock for this cell line~\citep{TFS16biosystems}. 
The cell line was modified
to include three fluorescent markers of the circadian clock and the cell cycle:
the RevErb-$\alpha$::Venus clock gene reporter \citep{NSB04cell} for measuring
the expression of the circadian protein RevErb$\alpha$, and the Fluorescence
Ubiquitination Cell Cycle Indicators (FUCCI), Cdt1 and Geminin, two cell cycle
proteins which accumulate during the G1 and S/G2/M phases respectively,
for measuring the cell cycle phases~\citep{SKM08cell}. The cells were left to
proliferate \textit{in vitro} in standard culture medium supplemented with 20\%
of Fetal Bovine Serum. 
Fluorescence recording was performed in constant
conditions with one image taken every 15 to 30 minutes during 72 hours.

\begin{figure}[ht!]
    \includegraphics[width=1\linewidth]{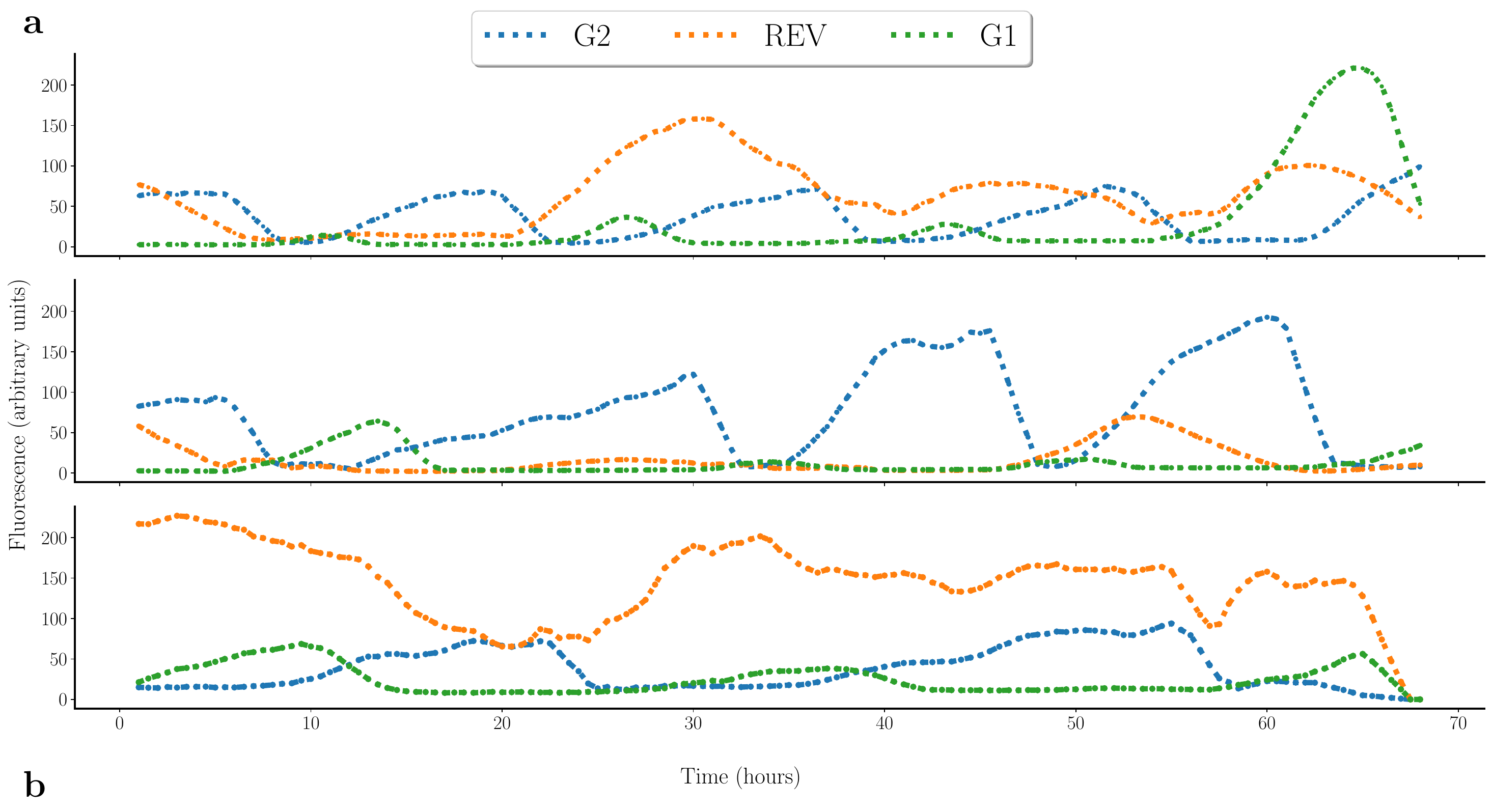}
    \includegraphics[width=1\linewidth]{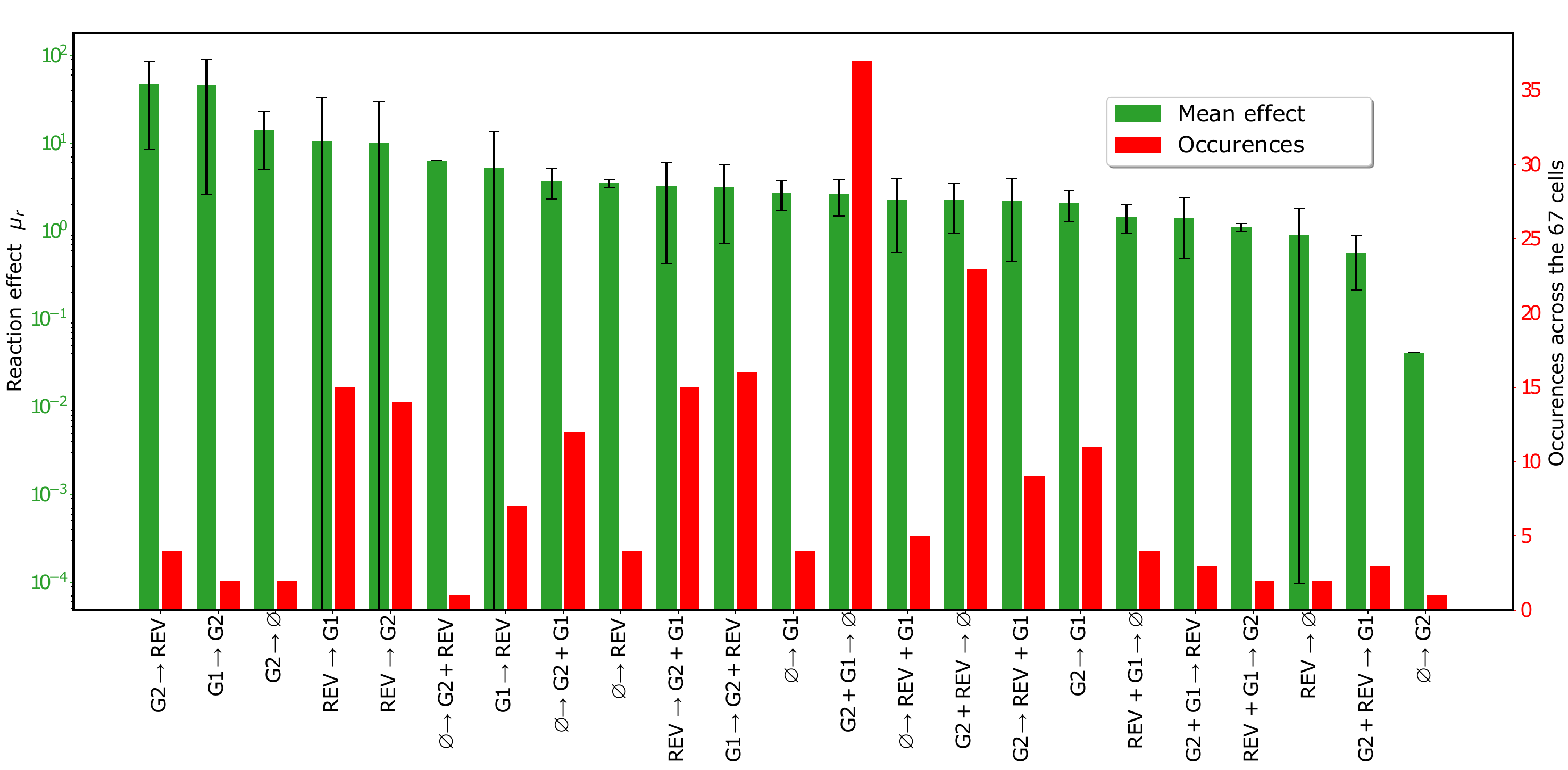}
    \caption{\textbf{Results of Reactmine on fluorescence videomicroscopy dataset. a} Plot of 3 cells among the 67 cells of the videomicroscopy data. Traces
    have been smoothed with a moving average. \textbf{b} Ranking of mean effect (green) and occurences (red) of inferred reactions on 67 cells.
    Bars in green report the mean effect of each reaction across
    time and cells in the videomicroscopy data.
    Standard deviation is computed across the cells.}
    \label{fig:fucci}
\end{figure}

A dataset of $67$ tracked cells was built from these experiments \citep{FKT14pnas},
giving rise to fluorescence levels trajectories with high inter-cell variability
and noise (Fig.~\ref{fig:fucci}\textbf{a}).
Thus, 
our learning protocol will 
\begin{itemize}
\item
first smooth the curves with a sliding window of 2.5 h;
\item
apply Reactmine to infer a CRN for each cell individually, using 
Michaelis-Menten rate functions (as described in Section~\ref{subsec:kinetic}) in order to fit indirect effects,
$\gamma=3$ in order to discover the main influences between the 3 variables,
and the remaining three hyperparameters selected by grid search as mentioned in Section~\ref{subsec:polish};
\item 
then compute statistics on the number of reaction occurrences across the $C=67$ inferred CRNs (one per cell). Let $y_i^{(c)}$ be the fluorescence vector at time $t_i$ for cell $c$.
We define the mean effect $\mu$ of a reaction $r = (R, P, f)$ on the derivative matrix as:
\begin{equation}
    \mu_r = \frac{1}{n C} \sum_{c=1}^{C} \sum_{i=1}^n f(y_i^{(c)})
\label{eq:effectfucci}
\end{equation}
\end{itemize}

Two reactions clearly stand out compared to the others in terms of effect (Fig.~\ref{fig:fucci}\textbf{b}).
The first one , $G2 \reac{} REV$,
represents a possible effect of the cell cycle on the circadian clock through the activation of  Rev-Erb$\alpha$ during the G2 phase,
in agreement with the main surprising findings of the modeling study of \citep{TFS16biosystems}, using the same
dataset.
The second most impactful reaction effect-wise
is $G1 \reac{} G2$, the reaction accounting for the cell phase transition from G1 into
S/G2/M. 
The reverse formal reaction $G1 \reac{} G2$ which could perhaps be expected,
is learned but ranked effect-wise behind several meaningless reactions.
In terms of occurrence number, one can observe the
predominance of a meaningless reaction $G2+G1\reac{} \emptyset$, present
in 37 out of 67 cells, yet never inferred first, and with low rate constants.
 
\subsection{Detection of systemic controls on clock gene transcription}

\begin{figure*}[ht!]
\centering
    \includegraphics[width=.85\linewidth]{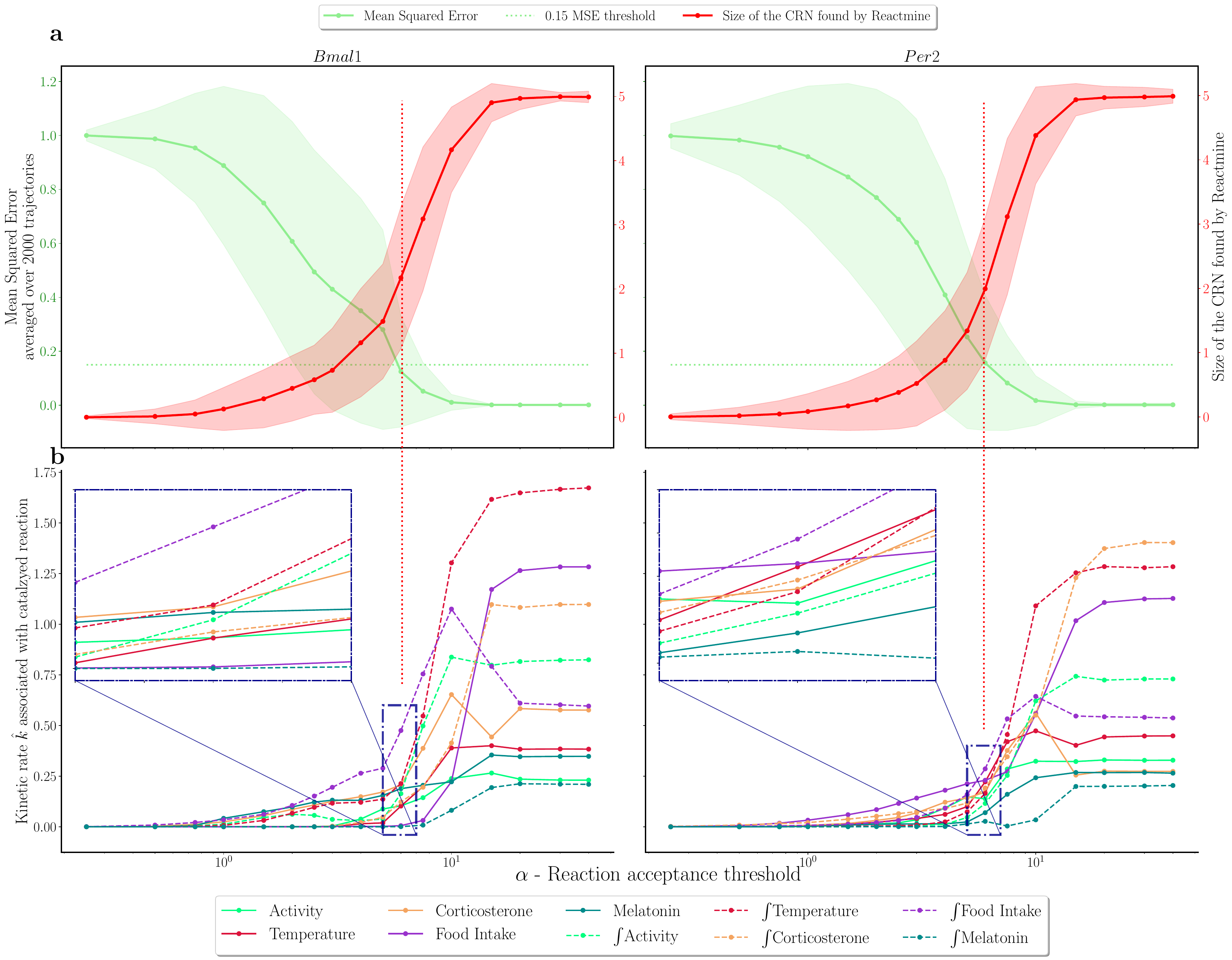}
    \caption{
    \textbf{Inference of systemic drivers for \emph{Bmal1} and \emph{Per2} clock genes transcription by Reactmine. a}
    Average and std for the mean square error (green) and number of reactions (red) for the CRNs
    inferred by Reactmine, as a function of $\alpha$ the reaction acceptance threshold.
    The statistics are computed over the $N=2000$
    trajectories $\{y^{(j)}\}_{1\le j \le N}$ considered.
    The green dotted horizontal line represents a MSE threshold
    of 0.15, equivalent here to 15\% of the variance left unexplained by the inferred CRN.
    The red dotted vertical line reveals that for both \emph{Bmal1} and
    \emph{Per2}, CRNs with only two reactions are enough to reach a MSE below 0.15.
    \textbf{b}
    Mean rate constant associated to a catalyzed reaction
    reported as a function of $\alpha$. The mean is computed over
    the $N=2000$ trajectories $\{y^{(j)}\}_{1\le j \le N}$.
    The plot zoom corresponds to $\alpha\approx6$, equivalent to an average number of two reactions per CRN.}
    \label{fig:eccb}
\end{figure*}

In this section, we 
apply Reactmine to biological data measuring
the circadian rhythms of five systemic regulators in mice: body temperature, rest-activity
rhythms, food intake, plasma corticosterone and melatonin, as well as the circadian
mRNA expression data in the mouse liver of two core clock genes: \emph{Bmal1} and \emph{Per2}.
These datasets have been analyzed in~\citep{MDL21bioinformatics}
to infer the preponderant systemic regulators on clock gene transcription
through another model learning approach. Using an ODE-based model of the
liver circadian clock~\citep{HMA21csbj} as well as data from four mouse classes,
transcription activation profiles $y$ were derived for each gene.
These profiles were regressed on systemic regulators
with the aim to infer the significant drivers of clock gene transcription. 
Let us now use Reactmine for this task and compare the results.

Since we are looking for an influence model rather than a reaction model properly speaking, 
we shall use the classical encoding of an influence by one formal reaction catalyzed by the source of the influence~\citep{FMRS18tcbb}.
We will thus enforce in Reactmine the search of influence reactions of the form
$z \reac{k} z + y$ where $z$ is one systemic regulator, $y$ one transcription activation profile for a target gene.
We shall assume mass action law kinetics representing influence forces.

Furthermore, it is assumed that a systemic regulator $z$ acts either directly, or indirectly though intermediate species by considering its integral counterpart $\int z$, but not both ways.
The 5 systemic regulators thus lead to 10 possible influences on \emph{Bmal1} and \emph{Per2} genes.
Reactmine hyperparameters $\gamma$ and $\beta$ can be set accordingly to 5 and 10.
$\delta_{\max}=3$ as in previous experiments.
Only  $\alpha$, the influence acceptance threshold, needs to be searched in order to restrict ourselves to preponderant influences only.
Fig.~\ref{fig:eccb}\textbf{a} shows the mean quadratic loss and number
of influences of the CRNs inferred by Reactmine for different values of $\alpha$.
For $\alpha=6$, the inferred CRNs with only two influences in average
reach an MSE below 0.15 - or equivalently - explain more than $85\%$ of the variance.

\Cref{fig:eccb}\textbf{b} recapitulates the mean kinetic rate constants of the
inferred reactions obtained across the traces as a function of $\alpha$, with the zoom part
corresponding to $\alpha=6$,  for which the inferred CRNs contain 2 reactions in average.
Remarkably, Reactmine  discovers that Food Intake and Temperature are the main
influencing factors of clock gene transcription, either
in a direct or an indirect manner.
{Using $\alpha=6$}, for \emph{Bmal1},
the indirect action of Food Intake is deemed the most relevant regulator, 
followed by the indirect action of Temperature and the direct action of Corticosterone, in agreement with~\cite{MDL21bioinformatics}.
Concerning \emph{Per2}, Food Intake and Temperature again stand out as most important systemic drivers. 
The recovered regulation importance ordering is the same as in~\citep{MDL21bioinformatics} (Fig. 6B-C),
except for one permutation between $\int$Corticosterone and $\int$Temperature.

\section{Conclusion}

Reactmine is an algorithm specially designed to infer biochemical reactions with kinetics,
between molecular species observed in \emph{wild type} time series data, i.e.\ without
gene knock out nor other possibilities to put initial conditions to $0$ at will.
On a benchmark of hidden CRNs of increasing difficulty, including one model of
MAPK signal transduction levels~\citep{QNK07ploscb},
we have shown that Reactmine is able to recover the hidden CRN, or an essentially equivalent
form of it, from one single ODE simulation trace. On the opposite,
the state-of-the-art sparse regression algorithm for non linear dynamical systems,
SINDy~\cite{BPK16pnas}, with appropriate function libraries for the examples,
fails to infer sparse ODE systems in that setting,
and even to reproduce the simulation traces obtained from different initial states.

The behavior of sparse regression algorithms is indeed conditional to assumptions about the
low level of correlation between predictors~\citep{ZY06jmlr}.
Those hypotheses are not satisfied in the context of \emph{wild type} time series data,
specifically in a low data regime.
The possibility of varying the traces by setting to 0 some initial conditions has
a de-correlation effect 
which may explain the better results reported in~\cite{MBP16ieee},
similarly to what has been shown for boolean models in~\cite{CFS17cmsb}.
Those arguments are the subject of current work in the general context
of sparse identification of non linear dynamics.

Reactmine solves those issues by not relying on sparse regression 
but on a statistical search algorithm with a bounded depth of reaction candidates which ensures sparsity by construction.
Reactmine uses four hyperparameters.
In all our examples, the maximum ratio $\delta_{\max}$ of variation between reactants and products for the inference of one reaction skeleton at a transition could be fixed to 3.
On the other hand, sensitivity analysis shows that the other three hyperparameters are important
and are currently determined by gridsearch in absence of better understanding of their sensitivty to the dimension and other characteristics of the time series data.
When applied to real biological data of videomicroscopy data
and systemic circadian controls of clock genes, we have shown that Reactmine is able to retrieve the main conclusions
of the model-based analyses done respectively in~\cite{TFS16biosystems} and~\cite{MDL21bioinformatics} on the same datasets.

These encouraging results should motivate applying Reactmine to new study cases on the one hand,
possibly dropping the restriction of 0/1 stoichiometry for a limited number of reaction schemas,
and on the other hand, generalizing Reactmine by introducing hidden variables in inconsistent transitions
to deal with the further challenge of learning models with latent species.

\bibliographystyle{natbib}
\bibliography{circadian}

\begin{thebibliography}{}

\bibitem[Aalto {\em et~al.}(2020)Aalto, Viitasaari, Ilmonen, Mombaerts, and
  Gonçalves]{AVI20nc}
Aalto, A., Viitasaari, L., Ilmonen, P., Mombaerts, L., and Gonçalves, J.
  (2020).
\newblock Gene regulatory network inference from sparsely sampled noisy data.
\newblock {\em Nature Communications\/}, {\bf 11}(1).

\bibitem[Brunton {\em et~al.}(2016)Brunton, Proctor, and Kutz]{BPK16pnas}
Brunton, S.~L., Proctor, J.~L., and Kutz, J.~N. (2016).
\newblock Discovering governing equations from data by sparse identification of
  nonlinear dynamical systems.
\newblock {\em Proceedings of the National Academy of Sciences\/}, pages
  3932--3937.

\bibitem[Carcano {\em et~al.}(2017)Carcano, Fages, and Soliman]{CFS17cmsb}
Carcano, A., Fages, F., and Soliman, S. (2017).
\newblock {Probably Approximately Correct Learning of Regulatory Networks from
  Time-Series Data}.
\newblock In {\em {CMSB'17}: Proceedings of the fifteenth international
  conference on Computational Methods in Systems Biology\/}, volume 10545,
  pages 74--90.

\bibitem[Fages {\em et~al.}(2018)Fages, Martinez, Rosenblueth, and
  Soliman]{FMRS18tcbb}
Fages, F., Martinez, T., Rosenblueth, D., and Soliman, S. (2018).
\newblock Influence networks compared with reaction networks: Semantics,
  expressivity and attractors.
\newblock {\em IEEE/ACM Transactions on Computational Biology and
  Bioinformatics\/}.

\bibitem[Feillet {\em et~al.}(2014)Feillet, Krusche, Tamanini, Janssens,
  Downey, Martin, Teboul, Saito, L{\'e}vi, Bretschneider, van~der Horst,
  Delaunay, and Rand]{FKT14pnas}
Feillet, C., Krusche, P., Tamanini, F., Janssens, R.~C., Downey, M.~J., Martin,
  P., Teboul, M., Saito, S., L{\'e}vi, F., Bretschneider, T., van~der Horst, G.
  T.~J., Delaunay, F., and Rand, D.~A. (2014).
\newblock Phase locking and multiple oscillating attractors for the coupled
  mammalian clock and cell cycle.
\newblock {\em Proceedings of the National Academy of Sciences of the United
  States of America\/}, {\bf 111}(27), 9928--9833.

\bibitem[Hesse {\em et~al.}(2021)Hesse, Martinelli, Aboumanify, Ballesta, and
  Relógio]{HMA21csbj}
Hesse, J., Martinelli, J., Aboumanify, O., Ballesta, A., and Relógio, A.
  (2021).
\newblock A mathematical model of the circadian clock and drug pharmacology to
  optimize irinotecan administration timing in colorectal cancer.
\newblock {\em Computational and Structural biology\/}, {\bf 19}, 5170--5183.

\bibitem[Huynh-Thu and Geurts(2018)Huynh-Thu and Geurts]{HG18sr}
Huynh-Thu, V.~A. and Geurts, P. (2018).
\newblock {dynGENIE3}: dynamical {GENIE3} for the inference of gene networks
  from time series expression data.
\newblock {\em Scientific Reports\/}, {\bf 8}.

\bibitem[King {\em et~al.}(2004)King, Whelan, Jones, Reiser, Bryant, Muggleton,
  Kell, and Oliver]{KWJRBMKO04nature}
King, R.~D., Whelan, K.~E., Jones, F.~M., Reiser, P. G.~K., Bryant, C.~H.,
  Muggleton, S.~H., Kell, D.~B., and Oliver, S.~G. (2004).
\newblock Functional genomic hypothesis generation and experimentation by a
  robot scientist.
\newblock {\em Nature\/}, {\bf 427}(6971), 247--252.

\bibitem[Mangan {\em et~al.}(2016)Mangan, Brunton, Proctor, and
  Kutz]{MBP16ieee}
Mangan, N.~M., Brunton, S.~L., Proctor, J.~L., and Kutz, J.~N. (2016).
\newblock Inferring biological networks by sparse identification of nonlinear
  dynamics.
\newblock {\em IEEE Transactions on Molecular, Biological and Multi-Scale
  Communications\/}, {\bf 2}(1), 52--63.

\bibitem[Martinelli {\em et~al.}(2019)Martinelli, Grignard, Soliman, , and
  Fages]{MGSF19cmsb}
Martinelli, J., Grignard, J., Soliman, S., , and Fages, F. (2019).
\newblock On inferring reactions from data time series by a statistical
  learning greedy heuristics.
\newblock In {\em {CMSB'19}: Proceedings of the seventeenth Int. Conf. on
  Computational Methods in Systems Biology\/}, volume 11773 of {\em Lecture
  Notes in BioInformatics\/}. Springer-Verlag.

\bibitem[Martinelli {\em et~al.}(2021)Martinelli, Dulong, Li, Teboul, Soliman,
  Lévi, Fages, and Ballesta]{MDL21bioinformatics}
Martinelli, J., Dulong, S., Li, X.-M., Teboul, M., Soliman, S., Lévi, F.,
  Fages, F., and Ballesta, A. (2021).
\newblock Model learning to identify systemic regulators of the peripheral
  circadian clock.
\newblock {\em Bioinformatics\/}, {\bf 37}(1), i401--i409.

\bibitem[Nagoshi {\em et~al.}(2004)Nagoshi, Saini, Bauer, Laroche, Naef, and
  Schibler]{NSB04cell}
Nagoshi, E., Saini, C., Bauer, C., Laroche, T., Naef, F., and Schibler, U.
  (2004).
\newblock Circadian gene expression in individual fibroblasts: cell-autonomous
  and self-sustained oscillators pass time to daughter cells.
\newblock {\em Cell\/}, {\bf 119}, 693--705.

\bibitem[Qiao {\em et~al.}(2007)Qiao, Nachbar, Kevrekidis, and
  Shvartsman]{QNK07ploscb}
Qiao, L., Nachbar, R.~B., Kevrekidis, I.~G., and Shvartsman, S.~Y. (2007).
\newblock Bistability and oscillations in the huang-ferrell model of mapk
  signaling.
\newblock {\em PLOS Computational Biology\/}, {\bf 3}(9), 1--8.

\bibitem[Sakaue-Sawano {\em et~al.}(2008)Sakaue-Sawano, Kurokawa, Morimura,
  Hanyu, Hama, Osawa, Kashiwagi, Fukami, Miyata, Miyoshi, Imamura, Ogawa,
  Masai, and Miyawaki]{SKM08cell}
Sakaue-Sawano, A., Kurokawa, H., Morimura, T., Hanyu, A., Hama, H., Osawa, H.,
  Kashiwagi, S., Fukami, K., Miyata, T., Miyoshi, H., Imamura, T., Ogawa, M.,
  Masai, H., and Miyawaki, A. (2008).
\newblock {{V}isualizing spatiotemporal dynamics of multicellular cell-cycle
  progression}.
\newblock {\em Cell\/}, {\bf 132}(3), 487--498.

\bibitem[Stolovitzky {\em et~al.}(2007)Stolovitzky, Monroe, and
  Califano]{SMC07anyas}
Stolovitzky, G., Monroe, D., and Califano, A. (2007).
\newblock Dialogue on reverse-engineering assessment and methods: the {DREAM}
  of high-throughput pathway inference.
\newblock {\em Annals of the New York Academy of Sciences\/}, {\bf 1115}(1),
  1--22.

\bibitem[Traynard {\em et~al.}(2016)Traynard, Feillet, Soliman, Delaunay, and
  Fages]{TFS16biosystems}
Traynard, P., Feillet, C., Soliman, S., Delaunay, F., and Fages, F. (2016).
\newblock Model-based investigation of the circadian clock and cell cycle
  coupling in mouse embryonic fibroblasts: Prediction of reverb$\alpha$
  up-regulation during mitosis.
\newblock {\em Biosystems\/}, {\bf 149}, 59--69.

\bibitem[Zhao and Yu(2006)Zhao and Yu]{ZY06jmlr}
Zhao, P. and Yu, B. (2006).
\newblock On model selection consistency of lasso.
\newblock {\em Journal of Machine Learning Research\/}, {\bf 7}, 2541--2563.

\bibitem[Zoppoli {\em et~al.}(2010)Zoppoli, Morganella, and
  Ceccarelli]{ZMC10bmcb}
Zoppoli, P., Morganella, S., and Ceccarelli, M. (2010).
\newblock Timedelay-aracne: Reverse engineering of gene networks from
  time-course data by an information theoretic approach.
\newblock {\em BMC Bioinformatics\/}.

\end{thebibliography}

\newpage

\setcounter{figure}{0}
\setcounter{section}{0}
\setcounter{table}{0}

\renewcommand{\thefigure}{S\arabic{figure}}
\renewcommand{\thesection}{S\arabic{section}}
\renewcommand\thetable{S\arabic{table}}

\definecolor{colorf}{rgb}{0.5490196078431373, 0.33725490196078434, 0.29411764705882354}
\definecolor{colorg}{rgb}{0.8901960784313725, 0.4666666666666667, 0.7607843137254902}
\definecolor{colorh}{rgb}{0.4980392156862745, 0.4980392156862745, 0.4980392156862745}

\begin{LARGE}
\begin{center}
Supplementary Material for Reactmine: a statistical search algorithm for inferring chemical reactions from time series data
\end{center}
\end{LARGE}

\section{Extending Reactmine to Michaelis Menten and Hill Kinetics}\label{sec:suppmm}

Reactmine is also compatible with other forms of kinetics,
provided that a measure of quality of reaction parameters can be computed, such
as the coefficient of variation.
A single-reactant reaction $(R, P, f)$ follows Michaelis Menten kinetics
if $\forall j \in R\cup P,~ \forall i \in \{1, \dots, n\}$

\begin{equation}
    v_{i, j} = s_j f(\Y_{i, \bullet}) = s_j \nu_{\max} \frac{y_{i,u}}{y_{i,u} + K_m}
\end{equation}

\noindent where $\nu_{\max}$ and $K_m$ are parameters, and $R=\{u\}$.\@ Note that Michaelis Menten kinetics are not defined for bimolecular reactions.

As $y_u \to +\infty,~\lvert v_j\rvert \to \nu_{\max}$. Besides, $\lvert v_j\rvert$
being an increasing function of $y_u$, an estimator of $\nu_{\max}$
can be obtained as the highest observed derivative $\lvert v_{i, j}\rvert$ on the whole transitions: for all $j \in R\cup P$

\begin{equation}
    \tilde{\nu}_{\max, j} = \underset{i \in \{1, \dots, n\}}{\max} \lvert v_{i, j}\rvert
\end{equation}
Then, $K_m$ is defined as the value of reactant concentration for which the associated
derivative is equal to $\frac{\nu_{\max}}{2}$.
Since measurements are only available at discrete time points, one has

\begin{equation}
    \hat{K}_{m,j} = y_{i^*, u} \quad \text{ with } i^* =  \underset{i \in \{1, \dots, n\}}{\argmin}~\left\lvert \lvert v_{i,j}\rvert - \frac{\tilde{\nu}_{\max, j}}{2} \right\rvert
\end{equation}
Once an estimator for $K_m$ has been provided, we apply the same principle as in
Equation~\ref{eq:hatk} from the main text, to obtain a new estimator of $\nu_{\max}~\forall j \in R\cup P$

\begin{equation}
    \hat{\nu}_{\max, j} = \frac{s_j}{\#\mathcal{T}(r)} \sum_{i \in \mathcal{T}(r)} v_{i,j} \frac{\hat{K}_{m, j} + y_{i,u}}{y_{i,u}}
\end{equation}

The computation described above also applies to Hill Kinetics of order $\eta$, $\forall j \in R\cup P,~ \forall i \in \{1, \dots, n\}$

\begin{equation}
    v_{i, j} = s_j f(\Y_{i, \bullet}) = s_j \nu_{\max} \frac{y_{i,u}^\eta}{y_{i,u}^\eta + K_m}
\end{equation}

However, for Michaelis-Menten and Hill kinetics, the optimization of $\hat{K}_m$ constants leads
to a non-convex problem with no convergence guarantees.

\section{Detailed results of Reactmine and SINDy}\label{sec:suppsindy}

Computation times reported here have been obtained on a Macbook M1 2020 13" with 8 cores.
For Reactmine, the grid search was parallelized. 
For \href{https://pypi.org/project/pysindy/}{SINDy}~\citep{BPK16pnas} we used the \href{https://github.com/dynamicslab/pysindy}{\texttt{pysindy} library}
with STLSQ (sequential least square thresholding) optimizer, as we observed that it yielded better results
than Lasso or SR3.
We chose to report in Table~\ref{tab:results} and in the main text, the ODE system inferred by SINDy with a value of regularization hyperparameter $\lambda$ 
that leads to the quadratic loss  (which is not zero due to numerical integration errors) of the ground truth CRN recovered by Reactmine.

\begin{table}[ht!]
    \centering
    \resizebox{\textwidth}{!}{
\begin{tabular}{llll}
                \toprule
                Name & Hidden CRN & CRN inferred by Reactmine & ODE system inferred by SINDy\\

                \midrule
                Chain & \makecell[l]{$\textbf{A}\reac{1} B$\\$B\reac{1} C$\\
                $C\reac{1}D$\\$D\reac{1}E$}
                           &
                \makecell[l]{$\gt{D\xrightarrow[1.00041]{0.99869} E}$\\[8pt]$\gt{C\xrightarrow[1.00071]{0.99836} D}$\\[8pt]$\gt{B\xrightarrow[0.99753]{1.00102} C}$\\[8pt]$\gt{A\xrightarrow[1.00107]{1.00069}
                B}$}
                           &
                           \makecell[l]{$\left\{\begin{array}{l}
                            \begin{aligned}
                                \dot{A} &= \gt{-1.00 A}\\[-7pt]
                                \dot{B} &= \gt{1.00 A -1.00 B}\\[-7pt]
                                \dot{C} &= \gt{1.03B-1.03C}~ \ft{+0.01D-0.06AB}\\[-7pt]
                                \dot{D} &= \ft{0.33B-0.64DE}\\[-7pt]
                                \dot{E} &= \gt{1.00D}
                            \end{aligned}\end{array}\right.$}\\[20pt]
                \midrule
                Loop & \makecell[l]{$\textbf{A}\reac{1}B$\\$B\reac{1} C$\\$C\reac{1} D$\\$D\reac{1}
                E$\\$E\reac{1} \textbf{A}$}
                           &
                \makecell[l]{$\gt{A\xrightarrow[1.00005]{0.99958}B}$\\[8pt]$\gt{B\xrightarrow[0.99243]{0.99748} C}$\\[8pt]$\gt{C\xrightarrow[0.99156]{0.99772} D}$\\[8pt]$\gt{D\xrightarrow[0.92045]{0.99771}
                E}$\\[8pt]$\gt{E\xrightarrow[1.00007]{0.99758} A}$}
                &
                           \makecell[l]{$\left\{\begin{array}{l}
                            \begin{aligned}
                                \dot{A} &= \gt{-1.00 A + 1.03 E}~ \ft{-0.006D-0.07AE-0.06DE}\\[-7pt]
                                \dot{B} &= \gt{1.00 A -1.00 B}~ \ft{+0.004C + 0.001 AB - 0.211AC-0.092BC}\\[-7pt]
                                \dot{C} &= \gt{1.14 B -1.18 C}~ \ft{- 0.002 D -0.17 A B + 0.39CD}\\[-7pt]
                                \dot{D} &= \ft{0.35 B -0.35 E}\\[-7pt]
                                \dot{E} &= \ft{0.39 C +0.457 E-4.21AE}
                            \end{aligned}\end{array}\right.$}\\[20pt]
                \midrule
                \makecell[l]{Reactant\\Parallel} & \makecell[l]{$\textbf{A}+C\reac{3} B+C$\\$\textbf{D}\reac{2} C$\\$\textbf{E}\reac{1}C$}
                           &
                \makecell[l]{$\gt{D\xrightarrow[1.97698]{2.00109} C}$\\[8pt]$\gt{E\xrightarrow[0.00526]{1.00101} C}$\\[8pt]$\gt{A+C\xrightarrow[2.95790]{2.94380}B+C}$}
                &
                           \makecell[l]{$\left\{\begin{array}{l}
                            \begin{aligned}
                                \dot{A} &= \ft{-1.12 A -510000276790.32 C -0.87 A B + 1.03 A D} \\[-7pt]
                                &\quad+ \gt{510000276788.57 A C}~\ft{+ 510000276790.32 B C -0.04 D C}\\[-5pt]
                                \dot{B} &= \ft{1.12 A + 510000276790.24 C + 0.87 A B -1.04 A D}\\[-7pt]
                                &\quad\gt{-510000276788.49 A C}~\ft{ -510000276790.24 B C + 0.04 D C}\\[-5pt] 
                                \dot{C} &= \gt{-2.20 D}~\ft{+ 0.02 E -0.19 A D + 0.40 D E}\\[-7pt]
                                \dot{D} &= \ft{0.07 A}~\gt{+ 1.96 D + 1.02 E}~\ft{ - 0.06 D E}\\[-7pt]
                                \dot{E} &= \gt{-1.00E}
                            \end{aligned}\end{array}\right.$}\\[20pt]
                \midrule
                \makecell[l]{Product\\Parallel} & \makecell[l]{$\textbf{A}+\textbf{C}\reac{1} B+\textbf{C}$\\$\textbf{C}\reac{2} D$\\$\textbf{C}\reac{3} E$}
                           &
                \makecell[l]{$\gt{C\xrightarrow[3.12657]{2.93602} E}$\\[8pt]$\gt{C\xrightarrow[2.08438]{1.95735} D}$\\[8pt]$\gt{A+C\xrightarrow[1.04002]{0.96372} B+C}$}
                &
                           \makecell[l]{
                            $\left\{\begin{array}{l}
                                \begin{aligned}
                                    \dot{A} &= \ft{-0.93C}\\[-7pt]
                                    \dot{B} &= \ft{0.93C}\\[-7pt]
                                    \dot{C} &= \ft{68364.34 -84189.93 A + 32887.73 C + 14190.64 D}\\[-7pt]
                                    &\quad \ft{+ 21285.97 E -17066.849 A C -40273.85 A E + 990.87 C D}\\[-7pt]
                                    &\quad \ft{+ 990.83 C D + 1486.24 C E + 7919.89 D E}\\[-5pt]
                                    \dot{D} &= \gt{3.73 C}~\ft{-1.84 A C}\\[-7pt]
                                    \dot{E} &= \ft{65012.55 A-60751.24 C-77885.60 E -4258.48 A C}\\[-7pt]
                                    &\quad  \ft{-10214.65 C E -27068.38 D E}
                                \end{aligned}\end{array}\right.$}\\
                                \midrule
                MAPK & \makecell[l]{$\textbf{A}\reac{0.0045}Ap$\\$Ap+\textbf{B}\reac{1000} ApB$\\$ApB\reac{150} Ap+\textbf{B}$\\$ApB\reac{150}
Ap+Bp$\\$Ap+Bp\reac{1000} ApBp$\\$ApBp\reac{150}Ap+Bp$\\$ApBp\reac{150}Ap+Bpp$}
            &\makecell[l]{$\gt{A\xrightarrow[0.0045]{0.00449}Ap}$\\[8pt]
                          $\ft{Bp+Ap\xrightarrow[499.96702]{499.96702} Bpp+Ap}$\\[8pt]
                          $\at{B+Ap\xrightarrow[500.01476]{500.01476} Ap}$\\[8pt]
                          $\gt{ApB\xrightarrow[150.03923]{150.04220}Ap+Bp}$\\[8pt]
                          $\gt{Ap+B\xrightarrow[501.19119]{501.19113} ApB}$\\[8pt]
                          $\gt{ApB\xrightarrow[150.36767]{150.37064}Ap+B}$\\[8pt]
                          $\gt{Ap+Bp\xrightarrow[517.78581]{517.78303}ApBp}$\\[8pt]
                          $\gt{ApBp\xrightarrow[155.32578]{155.33508}Ap+Bp}$\\[8pt]
                          $\at{Ap+B\xrightarrow[500.18851]{500.18846}ApB+B}$\\[8pt]
                          }                                
                          &
            \makecell[l]{
                $\left\{\begin{array}{l}
                    \begin{aligned}
                        \dot{Bpp} &= \ft{11764.89 -9818.81 Bpp -21809.21 Ap -64774.82 ApBp -9881.63 Bp}\\[-5pt]
                        &\quad \ft{+ 109653.06 ApB -10102.57 B -23028.83 A + 23087.90 Bpp \times Ap}\\[-5pt]
                        &\quad \ft{+ 47383.24 Bpp \times ApBp + 0.01 Bpp \times Bp -94598.54 Bpp \times ApB}\\[-5pt]
                        &\quad \ft{+ 0.05 Bpp \times B + 24104.25 Bpp \times A + 58119.14 Ap \times Bp}\\[-5pt]
                        &\quad \ft{+ 117674.08 Ap \times B + 68314.42 ApBp \times Bp + 239788.36 ApBp \times B}\\[-5pt]
                        &\quad \ft{-171491.97 Bp \times ApB + 0.03 Bp \times B + 45027.82 Bp \times A + 118690.34 B \times A}\\[-5pt]
                        \dot{Ap} &= \ft{0.003 Bp -0.001 Bpp \times Bp -0.004 Bpp \times B -0.002 Bp \times B}\\[-5pt]
                        \dot{ApBp} &= \ft{-0.002 Bp + 0.001 Bpp \times Bp + 0.004 Bpp \times B + 0.002 Bp \times B}\\[-5pt]
                        \dot{Bp} &= \ft{-11345.814 + 9469.53 Bpp + 20253.98 Ap}~ \gt{+62939.32 ApBp}\\[-5pt]
                        &\quad \ft{+ 9525.97 Bp -105529.77 ApB + 9401.39 B + 22299.11 A}\\[-5pt]
                        &\quad \ft{-21772.25 Bpp \times Ap -45730.13 Bpp \times ApBp -0.01 Bpp \times Bp}\\[-5pt]
                        &\quad \ft{+ 91003.53 Bpp \times ApB -0.05 Bpp \times B -23476.51 Bpp \times A}\\[-5pt]
                        &\quad \gt{-56249.34 Ap \times Bp}~\ft{ + 996.55 Ap \times B -64537.43 ApBp \times Bp}\\[-5pt]
                        &\quad \ft{-113908.83 ApBp \times B + 163092.38 Bp \times ApB -0.03 Bp \times B}\\[-5pt]
                        &\quad \ft{-42276.50 Bp \times A + 113704.11 ApB \times B -763.549 B \times A}\\[-5pt]
                        \dot{ApB} &= 0\\[-5pt]
                        \dot{B} &= \ft{-668.78 + 557.83 Bpp + 1724.07 Ap + 2552.29 ApBp}\\[-5pt]
                        &\quad \ft{+ 564.79 Bp -5063.51 ApB + 551.20 B + 784.94 A}\\[-5pt]
                        &\quad \ft{-1609.99 Bpp \times Ap + -1960.37 Bpp \times ApBp -0.001 Bpp \times Bp}\\[-5pt]
                        &\quad \ft{+ 4399.19 Bpp \times ApB -0.01 Bpp \times B -827.38 Bpp \times A}\\[-5pt]
                        &\quad \ft{-3441.84 Ap \times Bp + 543.68 Ap \times B -4279.75 ApBp \times Bp}\\[-5pt]
                        &\quad \ft{-8537.02 ApBp \times B + 10869.45 Bp \times ApB}~\gt{-0.003 Bp \times B}\\[-5pt]
                        &\quad \ft{-3146.138 Bp \times A + 6610.523 ApB \times B + 1384.463 B \times A}\\
                        \dot{A} &= 0
                    \end{aligned}\end{array}\right.$
            }\\[20pt]
                \bottomrule
\end{tabular}
    }
\caption{\textbf{Results obtained by Reactmine and SINDy on hidden CRNs} using a single simulation trace from one initial state
containing the molecular species indicated in bold in the first column.
The learned reactions are indicated in green if they belong to the hidden CRN,
in yellow if they lead to equivalent terms of the associated ODEs, and in red otherwise. For a learned reaction,
the number written underneath the arrow is the initial rate constant value learned with the reaction, before global re-optimization.
For the ODE systems inferred by SINDy, the terms are coloured in green if they correspond to the kinetics of some hidden reactions, regardless of the precise kinetic constant value as long as the sign is exact, 
and in red otherwise.}
\label{tab:results}
\end{table}

\begin{table}[ht!]
    \centering
   \resizebox{\textwidth}{!}{
\begin{tabular}{llllll}
                \toprule
                CRN & \makecell[c]{Hyperparameters\\$(\alpha,\ \beta,\ \delta_{\max},~\gamma)$} & \makecell[c]{Computation time\\in seconds} & \makecell[l]{Number of\\inferred CRNs} &\makecell[c]{Grid search values for $(\alpha,\ \beta,~\delta_{\max},\ \gamma)$} & \makecell[c]{Computation time\\ with grid search}\\
                \midrule
                Chain &(0.02,\ 7,\ 3,\ 4) & 0.31 & 128 &\makecell[l]{$\alpha \in \{0.005, 0.01, 0.02, 0.05, 0.1, 0.15, 0.2, 0.25, 0.3, 0.5\}$\\$\beta \in \{4, 5, 6, 7, 8\}$\\$\delta_{\max}=3$\\$\gamma\in \{3,4,5,6\}$} & 3035.95\\

                \midrule
                Loop &(0.02,\ 7,\ 3,\ 6) & 40.78 & 4198 &\makecell[l]{$\alpha \in \{0.005, 0.01, 0.02, 0.05, 0.1, 0.15, 0.2, 0.25, 0.3, 0.5\}$\\$\beta \in \{4, 5, 6, 7, 8\}$\\$\delta_{\max}=3$\\$\gamma\in \{3,4,5,6\}$} & 30643.19\\
                \midrule
                Reactant-Parallel &(0.02,\ 4,\ 3,\ 5) & 1.81 & 265 &\makecell[l]{$\alpha \in \{0.005, 0.01, 0.02, 0.05, 0.1, 0.15, 0.2, 0.25, 0.3, 0.5\}$\\$\beta \in \{4, 5, 6, 7, 8\}$\\$\delta_{\max}=3$\\$\gamma\in \{3,4,5,6\}$} & 970.23\\
                \midrule
                Product-Parallel &(0.02,\ 5,\ 3,\ 3) & 0.13 & 24 &\makecell[l]{$\alpha \in \{0.005, 0.01, 0.02, 0.05, 0.1, 0.15, 0.2, 0.25, 0.3, 0.5\}$\\$\beta \in \{4, 5, 6, 7, 8\}$\\$\delta_{\max}=3$\\$\gamma\in \{3,4,5,6\}$} & 334.06\\
                \midrule
                MAPK & (0.02,\ 7,\ 3,\ 10)& 3012.85 & 33104 &\makecell[l]{$\alpha \in \{0.0025,0.005,0.0075,0.01,0.015,0.02,0.025,0.03,0.04,0.05,0.1,0.2\}$\\$\beta \in \{6, 8, 10\}$\\$\delta_{\max}=3$\\$\gamma\in \{7,8,9,10\}$} & 77179.92\\
                \bottomrule
\end{tabular}
   }
\caption{\textbf{Reactmine computation times and hyperparameter values used} for the results reported in Table~\ref{tab:results}.
The best hyperparameter settings indicated in the second column for each example, are found by grid search with a range of values reported in the third column, and a computation time given in the last column.
The CRN learning computation times in these settings are given in the third column, with the number of generated CRN candidates in the fourth column.}
\label{tab:computgrid}
\end{table}

\begin{figure}[ht!]
    \includegraphics[width=1\linewidth]{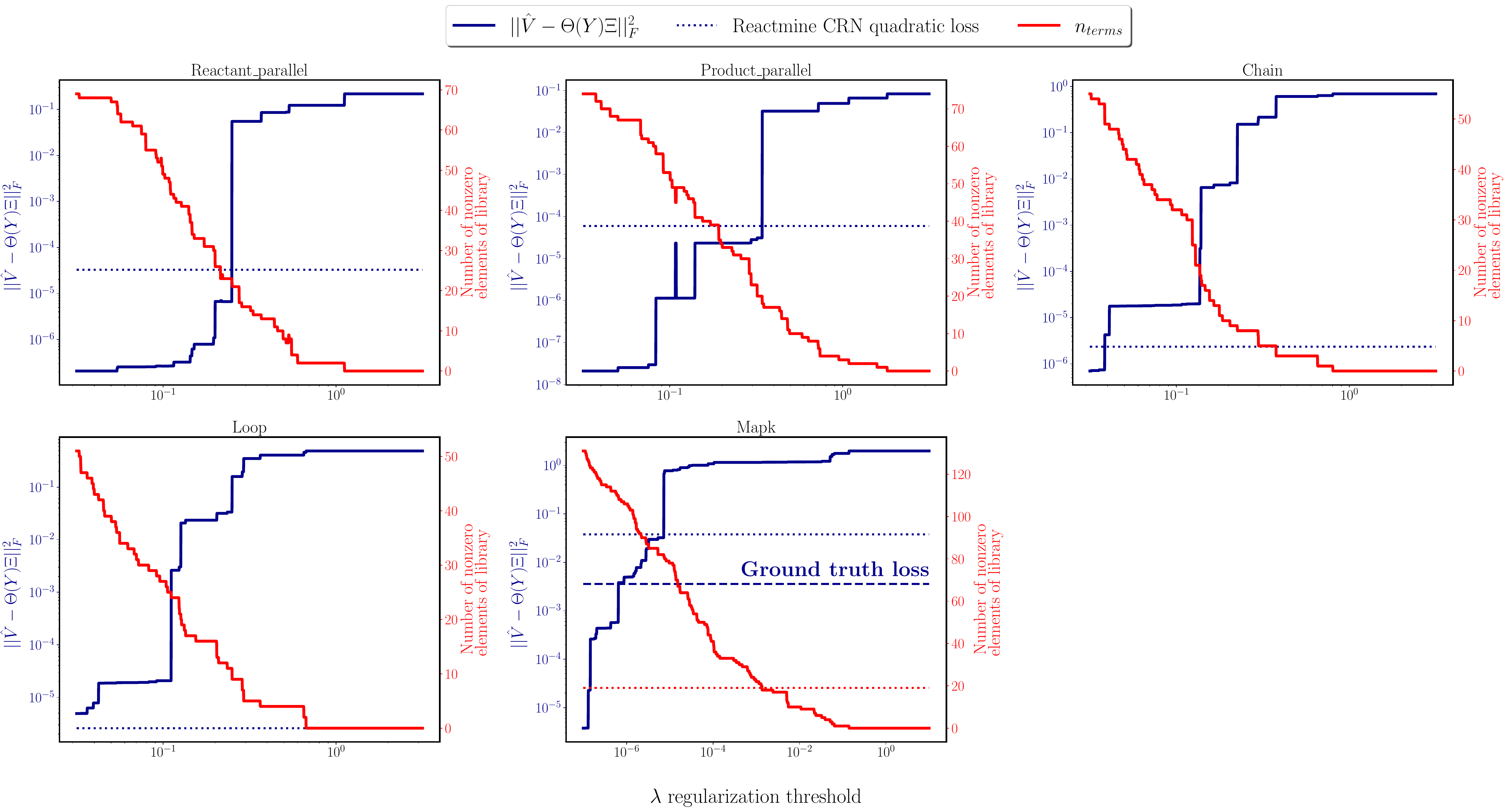}
    \caption{\textbf{Quadratic training loss (blue) and number of nonzero terms in the library (red) found by SINDy as a function of $\lambda$}
    in our benchmark of reactant-parallel, product-parallel, chain, loop and MAPK CRNs.
    The dashed red line represents the actual
    number of nonzero terms in the ground truth ODE associated to each hidden CRN. The dashed darkblue line stands for the (non-zero due to numerical errors) quadratic loss
    value of the ground truth CRN.}
    \label{fig:sindy}
\end{figure}

\begin{figure}[ht!]
    \includegraphics[width=1\linewidth]{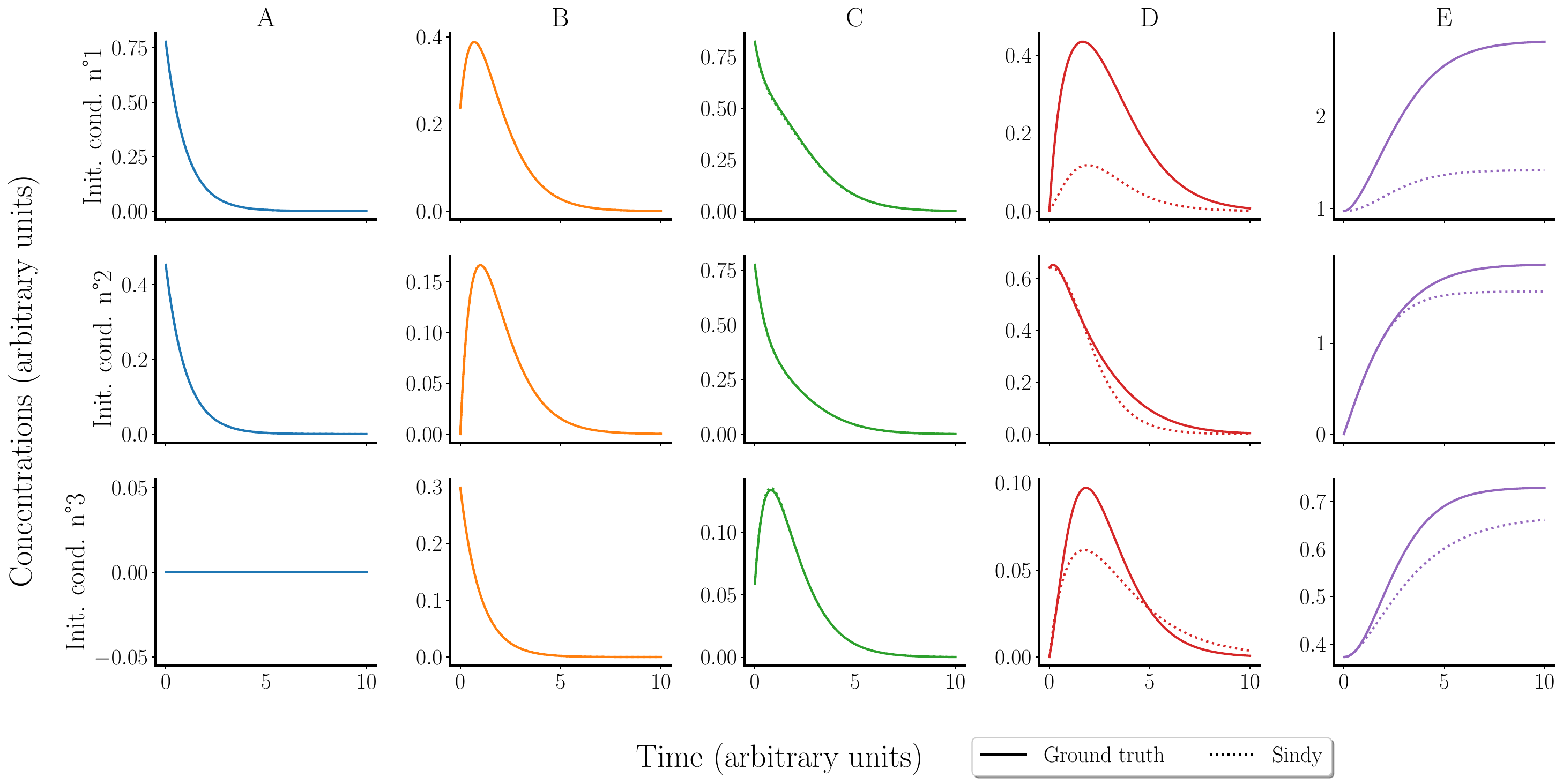}
\caption{\textbf{Simulation of the ODE model learned by SINDy in the Chain example using different initial conditions} (doted line) compared with ground truth (solid line),
showing erroneous dynamics for $D$ and $E$.}
\label{fig:sindy_nogen_chain}
\end{figure}

\begin{figure}[ht!]
    \centering
$\textcolor{colorg}{A}\reac{0.0045}
        \textcolor{colorb}{Ap}$\\
        $\textcolor{colorb}{Ap}+\textcolor{colorf}{B} \reac{1000} \textcolor{colore}{ApB}$\\
        $\textcolor{colore}{ApB} \reac{150}
    \textcolor{colorb}{Ap}+\textcolor{colorf}{B}$\\
    $\textcolor{colore}{ApB}\reac{150}\textcolor{colorb}{Ap}+\textcolor{colord}{Bp}$\\
    $\textcolor{colorb}{Ap}+\textcolor{colord}{Bp}\reac{1000}\textcolor{colorc}{ApBp}$\\
    $\textcolor{colorc}{ApBp}\reac{150}\textcolor{colorb}{Ap}+\textcolor{colord}{Bp}$\\
    $\textcolor{colorc}{ApBp}\reac{150}\textcolor{colorb}{Ap}+\textcolor{colora}{Bpp}$

    \includegraphics[width=1\linewidth]{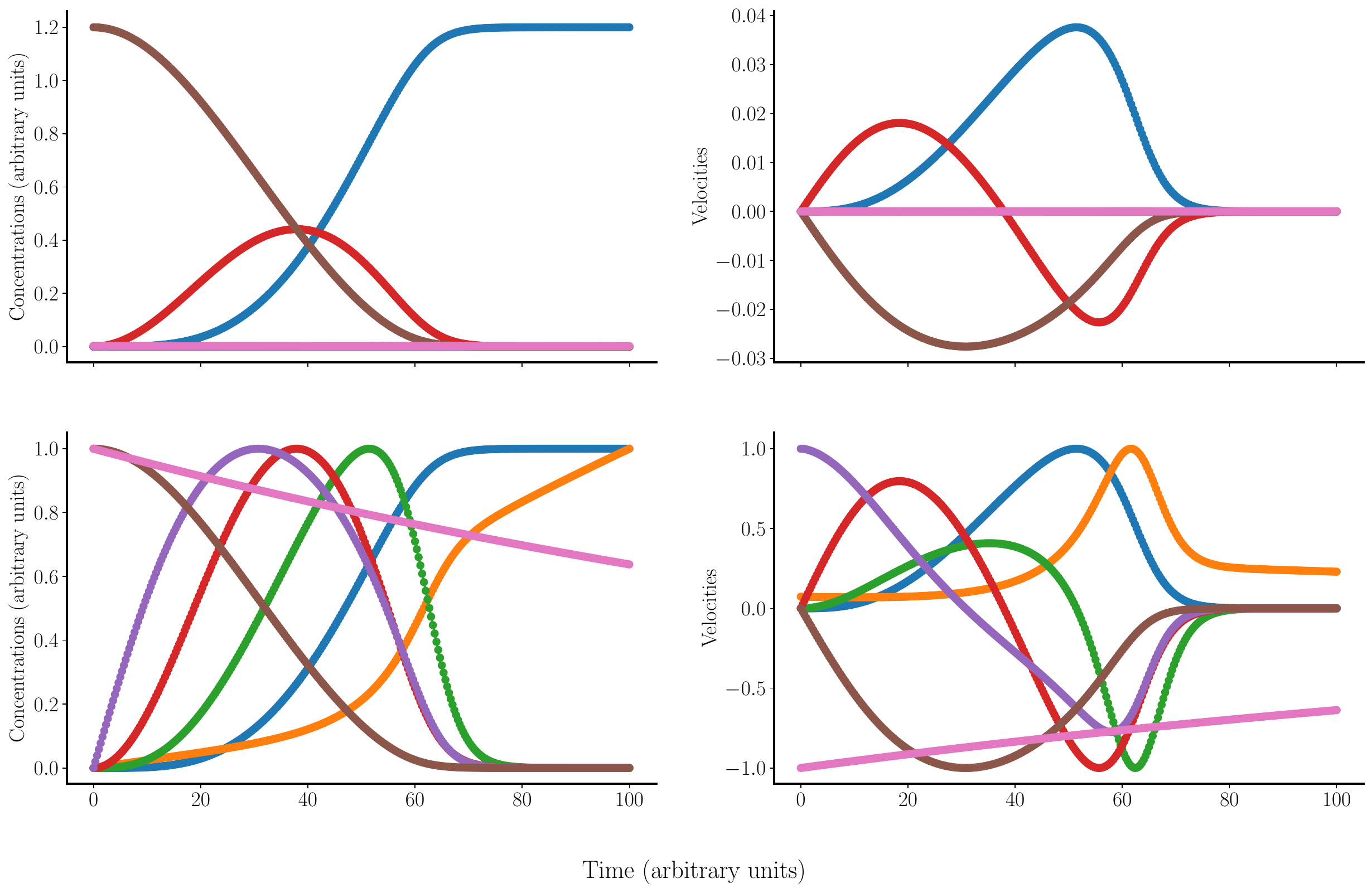}

        \caption{\textbf{Learning trace and estimated derivatives for the MAPK CRN.}
        The lower panel shows the same plots with
        concentrations and derivatives normalized by their maximal values along the trace
        for visualization purposes.}
        \label{fig:mapkplot}
        \end{figure}

\begin{figure}[ht!]
    \centering
    \includegraphics[width=1\linewidth]{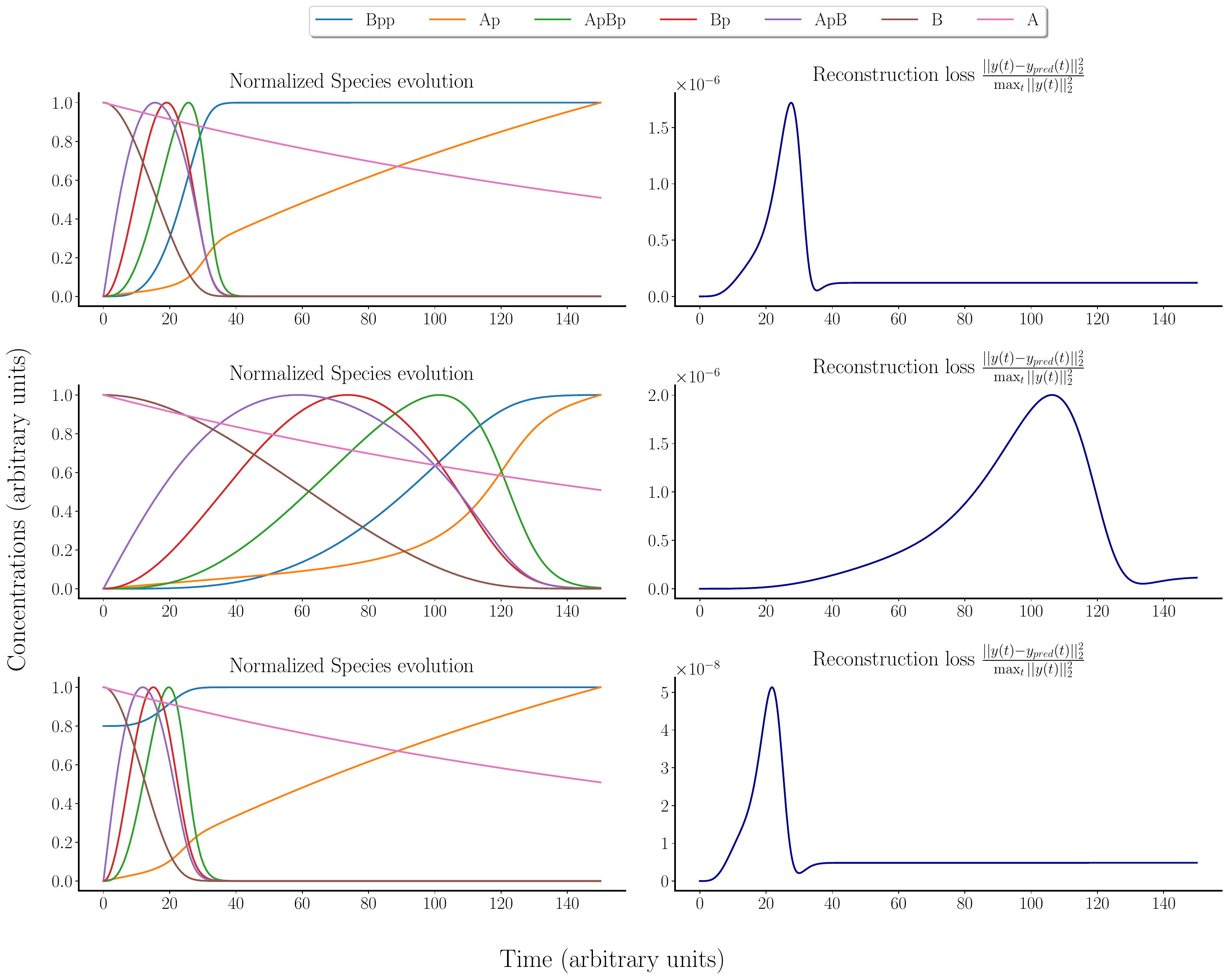}
    \caption{\textbf{Reproduction of the MAPK CRN simulation traces obtained from different initial conditions using the CRN inferred by Reactmine}.
    Time-resolved species concentrations are normalized to their maximal
    value across the trace. $y_0^{(1)}=(2,0,0,0,0,0.5,0.01);y_0^{(2)}=(0,0,0,0,0,1,0.1);
        y_0^{(3)}=(0,0,0,0,0,1.5,0.001)$. 
    The second column displays the loss between the simulation traces of the MAPK CRN and the CRN inferred by Reactmine.}
    \label{fig:mapk}
    \end{figure}

    \begin{figure}[ht!]
        \centering
        $\textcolor{colora}{A}+\textcolor{colorc}{C}\reac{3}
        \textcolor{colorb}{B}+\textcolor{colorc}{C}$\\
        $\textcolor{colord}{D} \reac{2} \textcolor{colorc}{C}$\\
        $\textcolor{colore}{E} \reac{1}
        \textcolor{colorc}{C}$
      
        \includegraphics[width=1\linewidth]{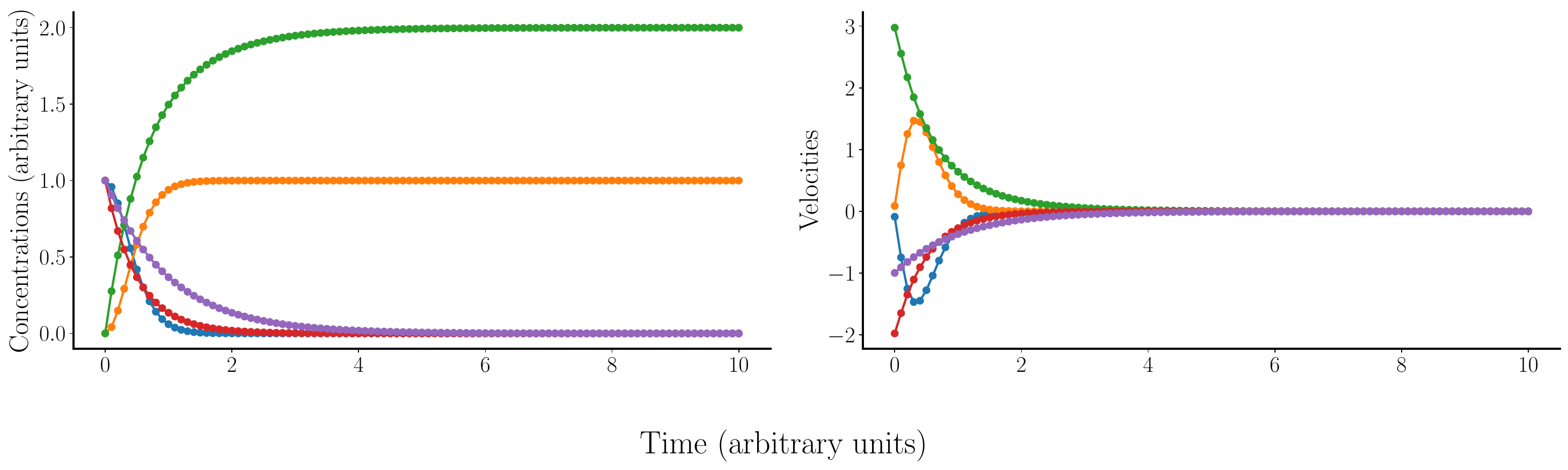}
        \caption{\textbf{Learning trace and estimated derivatives for the Reactant-Parallel CRN.}}
          \label{fig:reacplot}
      \end{figure}

      \begin{figure}[ht!]
        \centering
        $\textcolor{colora}{A}+\textcolor{colorc}{C}\reac{1}
        \textcolor{colorb}{B}+\textcolor{colorc}{C}$\\
        $\textcolor{colorc}{C} \reac{2} \textcolor{colord}{D}$\\
        $\textcolor{colorc}{C} \reac{3}
        \textcolor{colore}{E}$
        \includegraphics[width=1\linewidth]{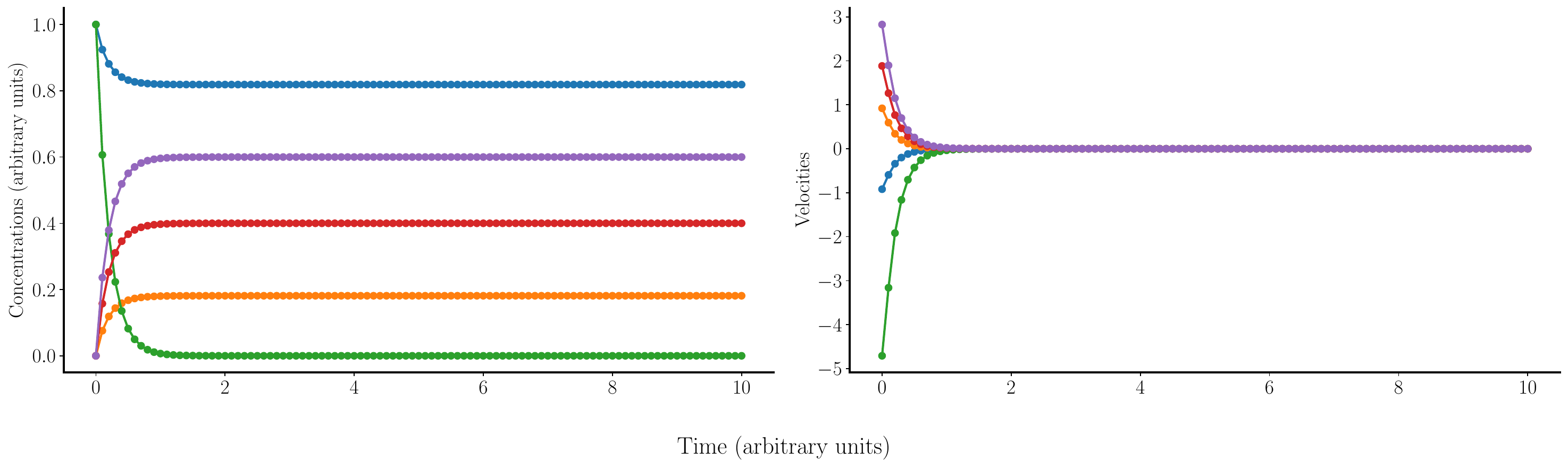}
        \caption{\textbf{Learning trace and estimated derivatives for the Product-Parallel CRN.}}
      \end{figure}
      
      \begin{figure}[ht!]
        \centering
        $\textcolor{colora}{A}\reac{1}
        \textcolor{colorb}{B} \reac{1} \textcolor{colorc}{C}
        \reac{1} \textcolor{colord}{D} \reac{1}\textcolor{colore}{E}$
      
        \includegraphics[width=1\linewidth]{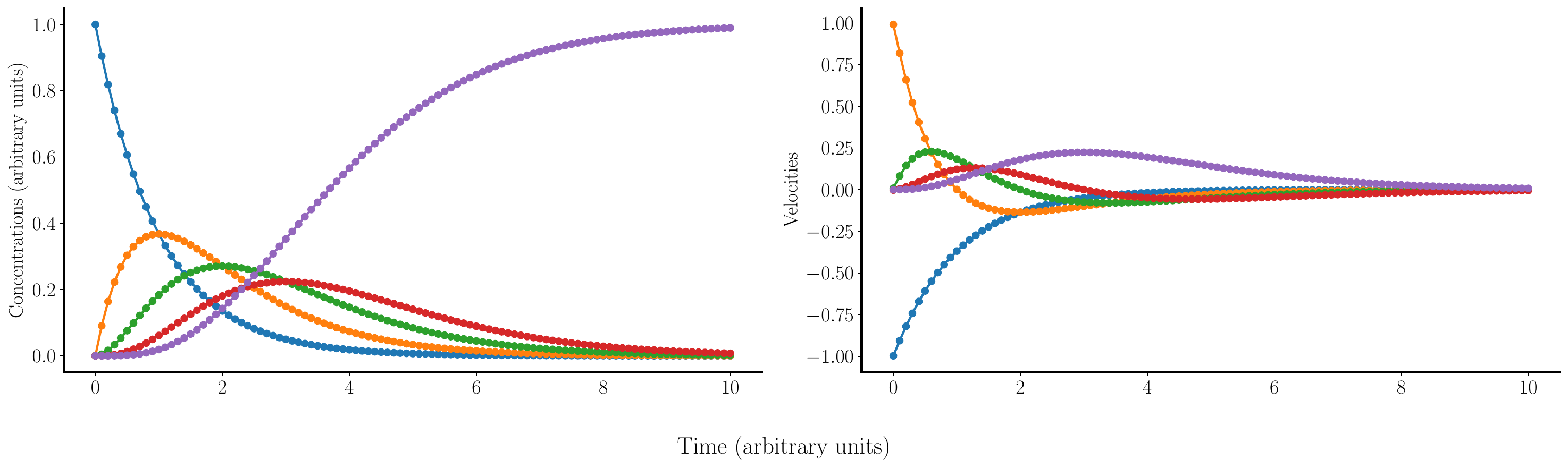}
        \caption{\textbf{Learning trace and estimated derivatives for the Chain CRN.}}
      \end{figure}
      
      \begin{figure}[ht!]
        \centering
        $\textcolor{colora}{A}\reac{1}
        \textcolor{colorb}{B} \reac{1} \textcolor{colorc}{C}
        \reac{1} \textcolor{colord}{D} \reac{1}
        \textcolor{colore}{E} \reac{1} \textcolor{colora}{A}$
      
        \includegraphics[width=1\linewidth]{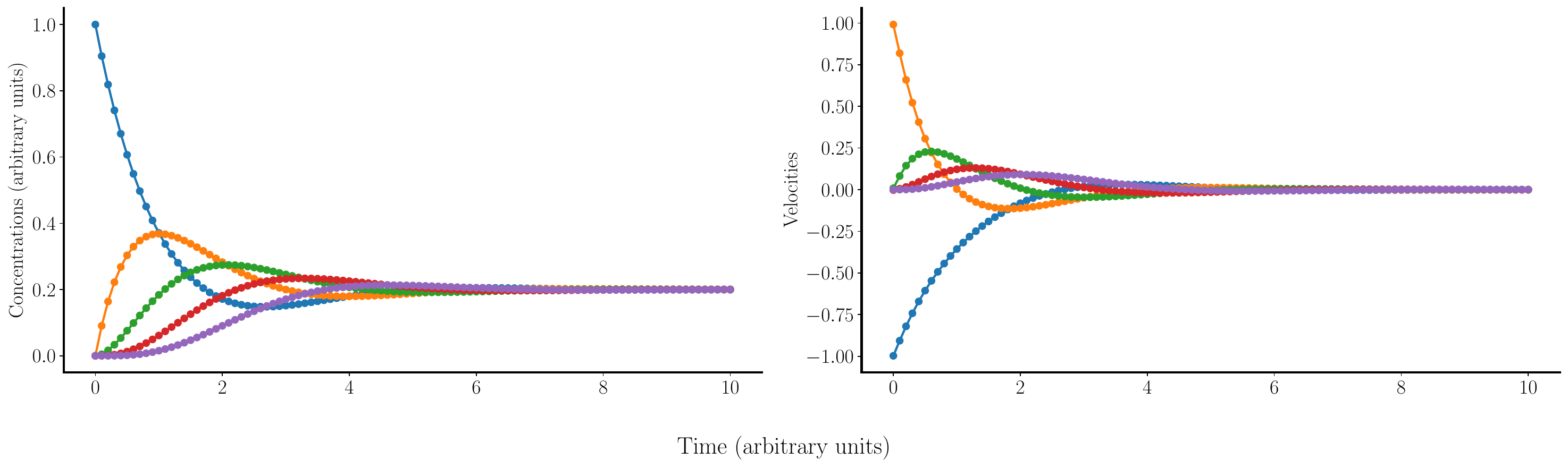}
        \caption{\textbf{Learning trace and estimated derivatives for the Loop CRN.}}
      \label{fig:loopplot}
      \end{figure}

\begin{figure}[ht!]
    \includegraphics[width=1\linewidth]{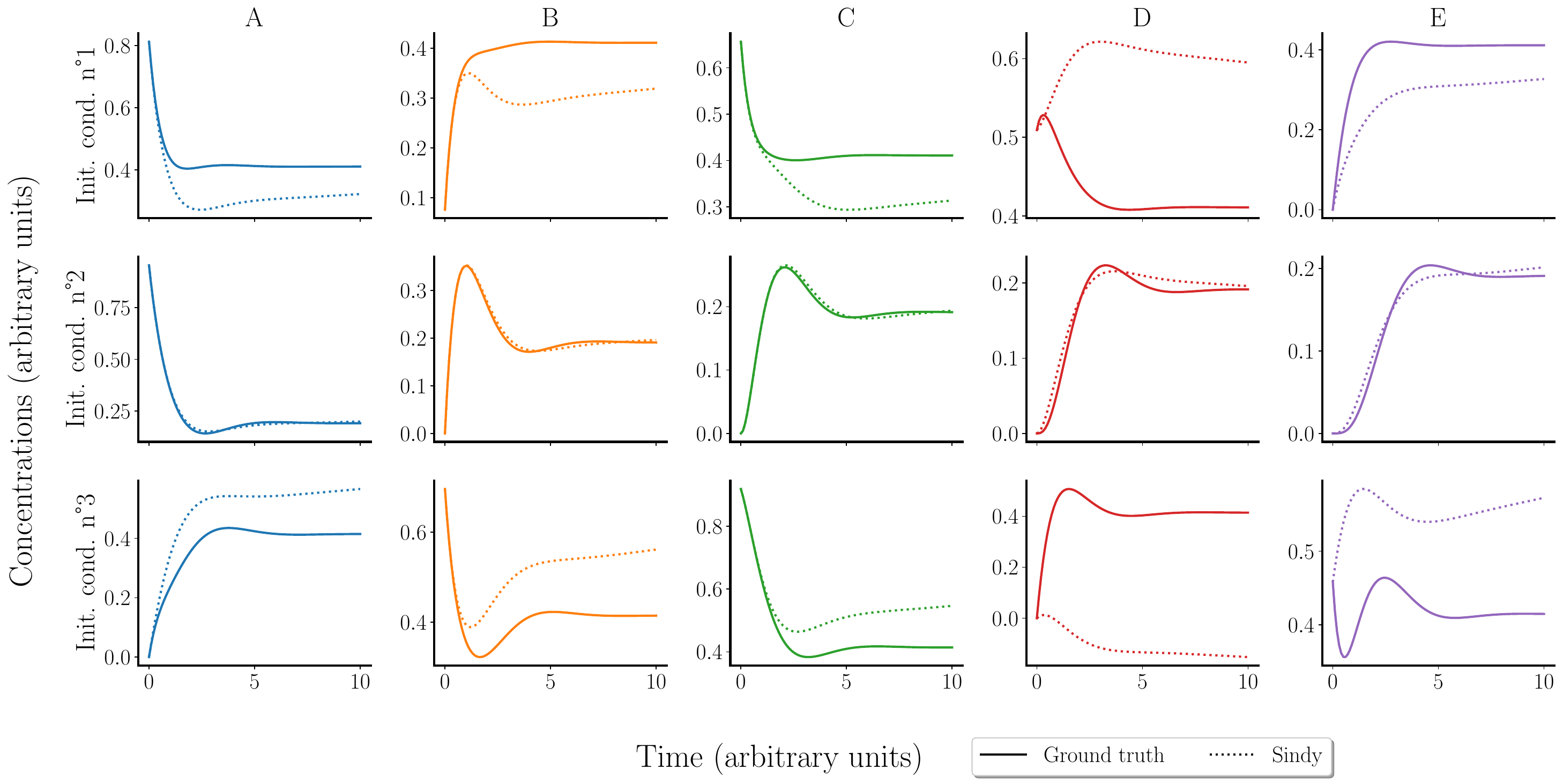}
  \caption{\textbf{Simulation of the ODE model learned by SINDy in the Loop example using different initial conditions} (doted line) compared with ground truth (solid line),
  showing erroneous dynamics on all species.}
\label{fig:sindy_nogen_loop}
\end{figure}

\begin{figure}[ht!]
    \centering
    \includegraphics[width=1\linewidth]{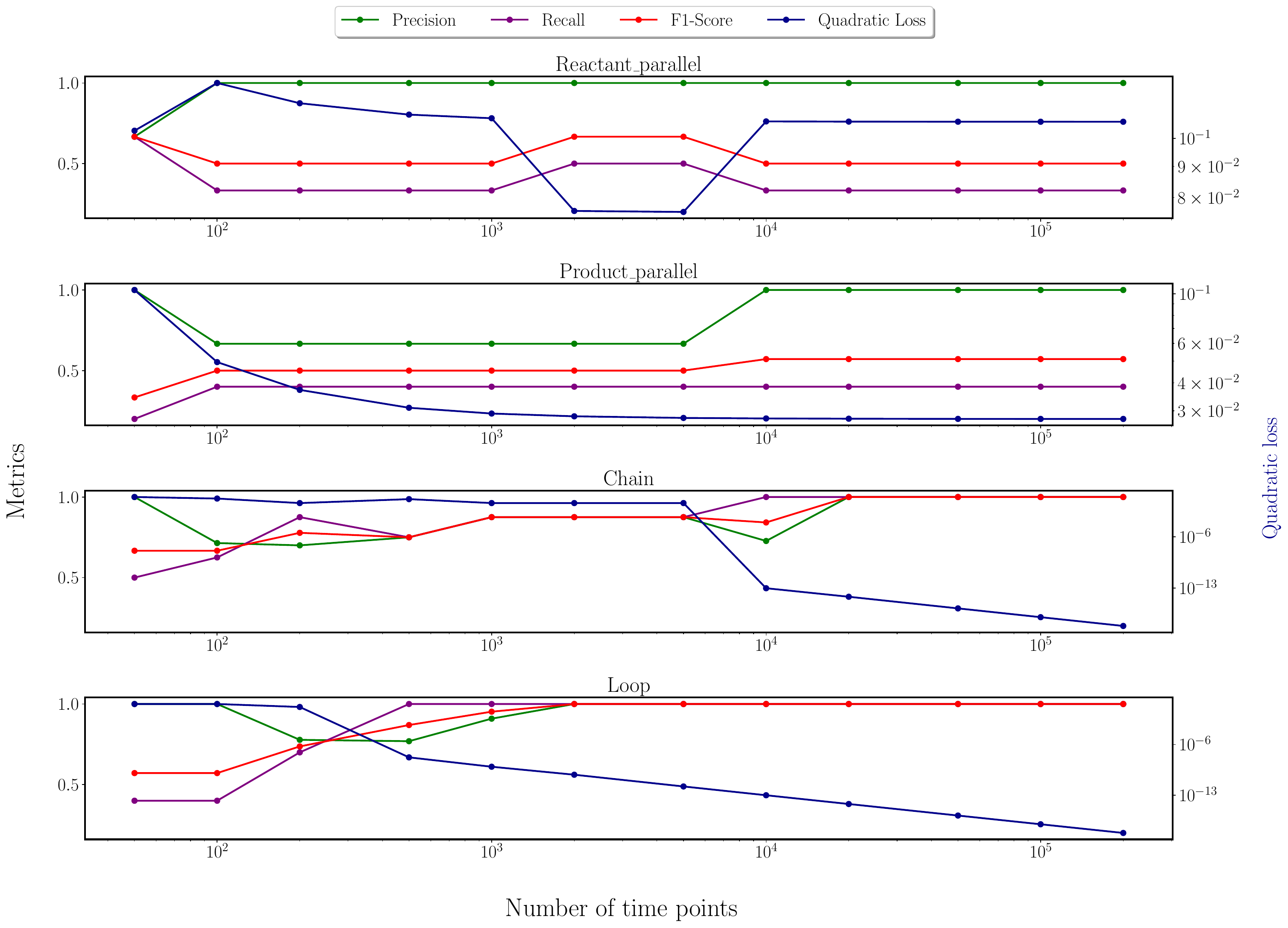}
    \caption{
        \textbf{Sensitivity of SINDy to the number of time points} with $\lambda$ selected to yield the highest binary $F_1$-Score 
        (i.e. $F_1$-score with labels 1 for terms with non-zero coefficient in the ODEs, independently of their sign, and labels 0 for zero coefficient).
     Precision, recall, F1-score  and quadratic loss are displayed as a function of the number of time points in the learning simulation trace used.
     With very high number of time points, SINDy succeeds in recovering the hidden chain and loop CRNs, 
     but this is not the case for the reactant-parallel and product-parallel CRNs.} \label{fig:sensitimechainSINDy}
\end{figure}

\begin{figure}[ht!]
    \centering
    \begin{subfigure}[t]{0.5\linewidth}
    \includegraphics[width=1\linewidth]{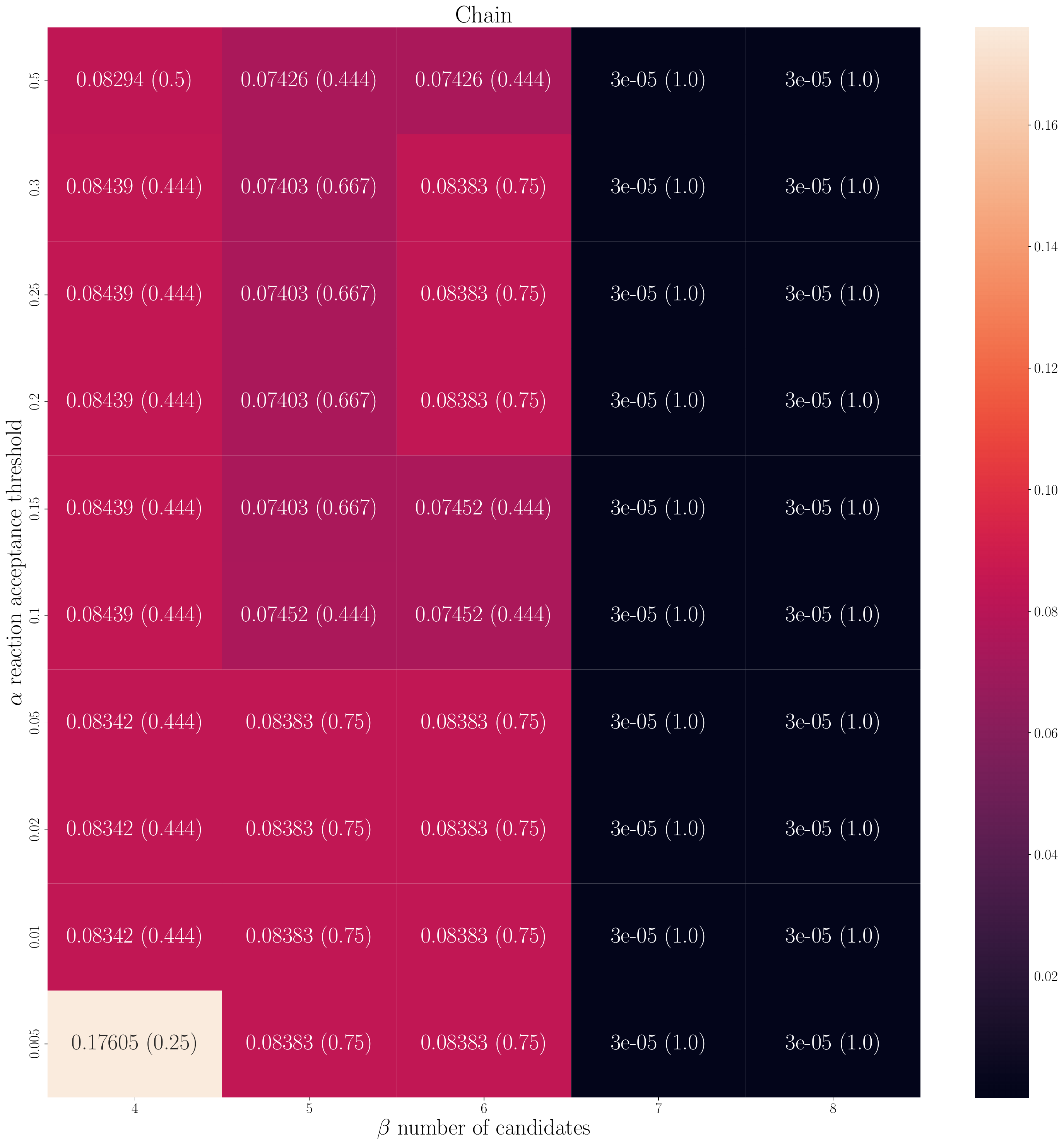}
    \caption{Chain CRN}
    \end{subfigure}%
    ~
    \vfill
    \begin{subfigure}[t]{0.5\linewidth}
        \includegraphics[width=1\linewidth]{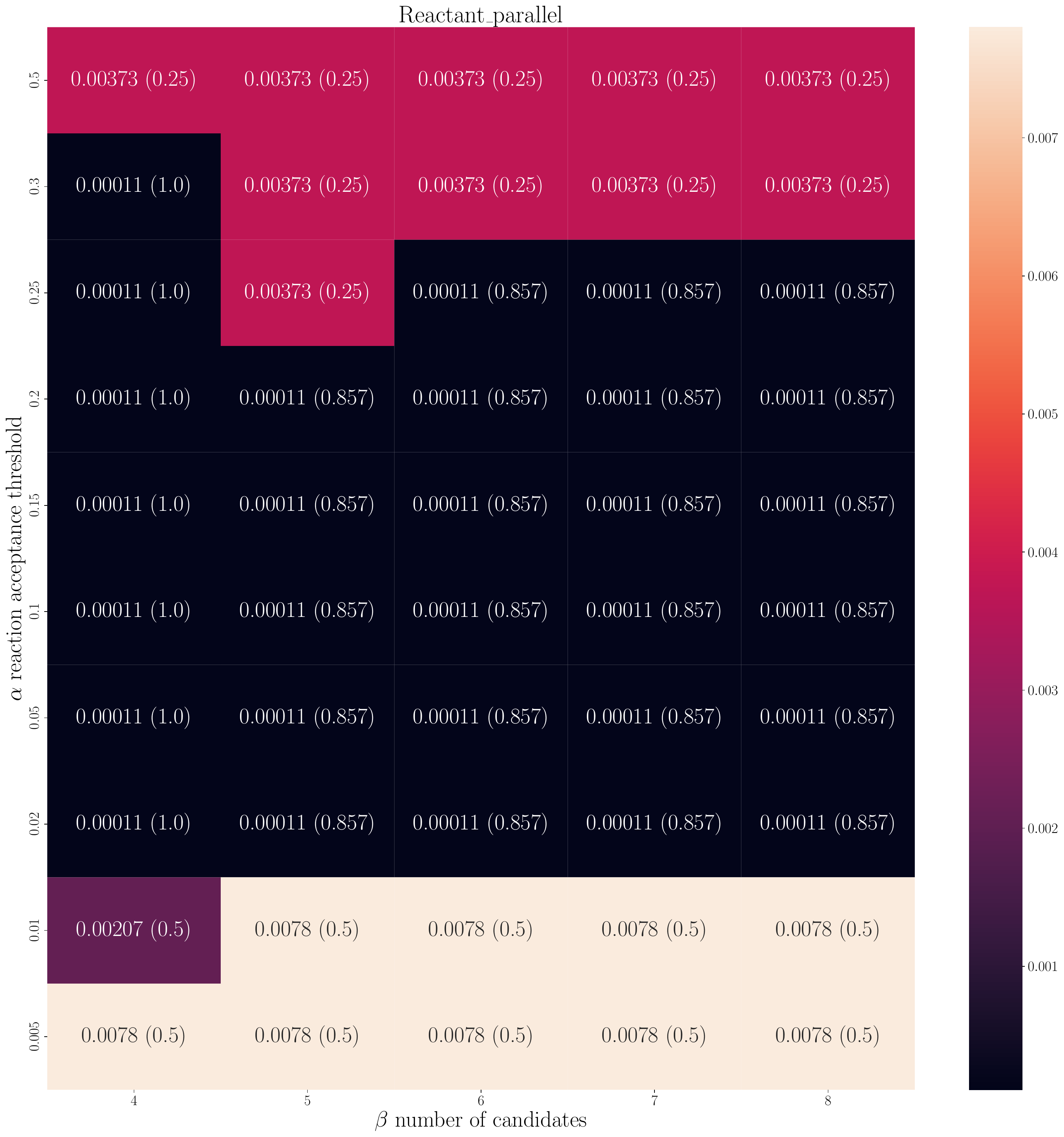}
        \caption{Reactant Parallel CRN}
    \end{subfigure}%
    ~
    \begin{subfigure}[t]{0.5\linewidth}
        \includegraphics[width=1\linewidth]{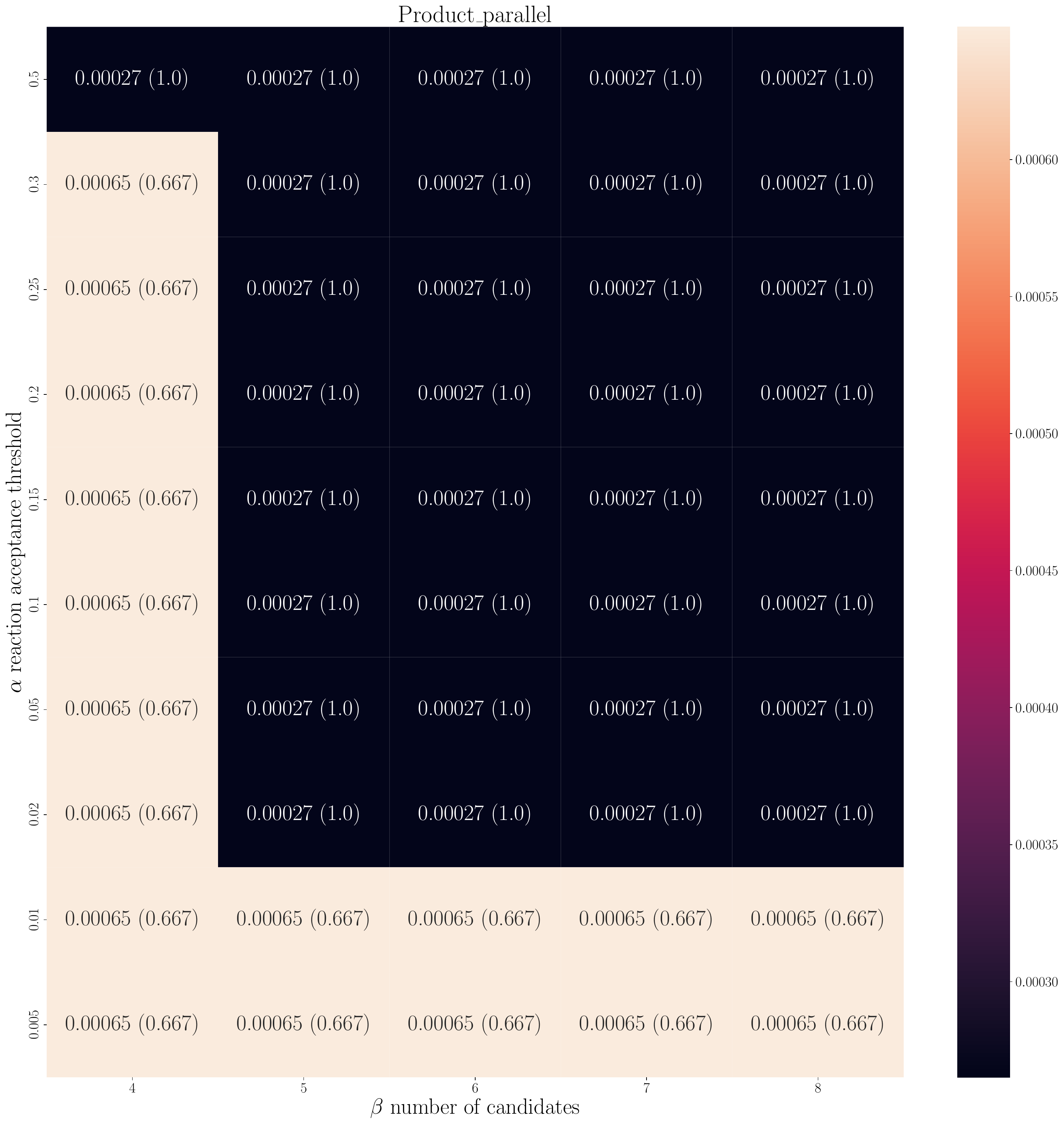}
        \caption{Product Parallel CRN}
    \end{subfigure}

    \caption{\textbf{Sensitivity of Reactmine to $\alpha$ and $\beta$ hyperparameters}.
    Quadratic loss $\norm{\V-\bF(\Y,\bk)\textbf{S}}_F^2$ (lower is better) and
     F1-score (in parenthesis, higher is better). The colorbar levels relate to the
     quadratic loss.
    $\delta_{\text{max}}=3$ and $\gamma=6$ except for the reactant-parallel CRN where $\gamma=5$,
    as it yielded a better result.}
    \label{fig:supheatmap}
 \end{figure}

\begin{figure}[ht!]
    \includegraphics[width=1\linewidth]{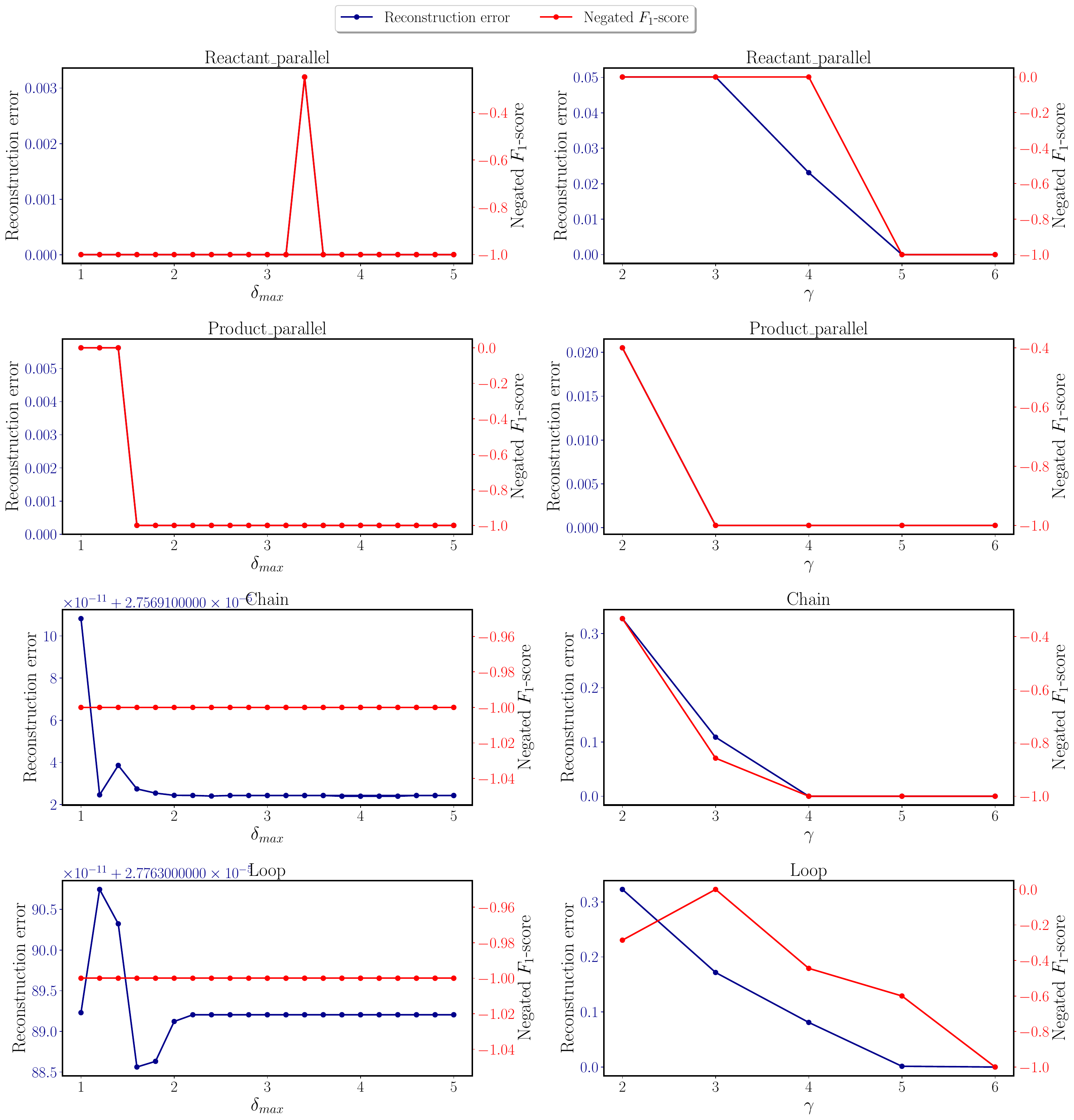}
    \caption{\textbf{Sensitivity of Reactmine to $\delta_{\max}$ and $\gamma$ hyperparameters}.
    The quadratic loss is reported in darkblue, the $F_1$-score in red, negated for 
    visualization purposes.
}
    \label{fig:sensideltagamma}
\end{figure}

\begin{figure}[ht!]
    \includegraphics[width=1\linewidth]{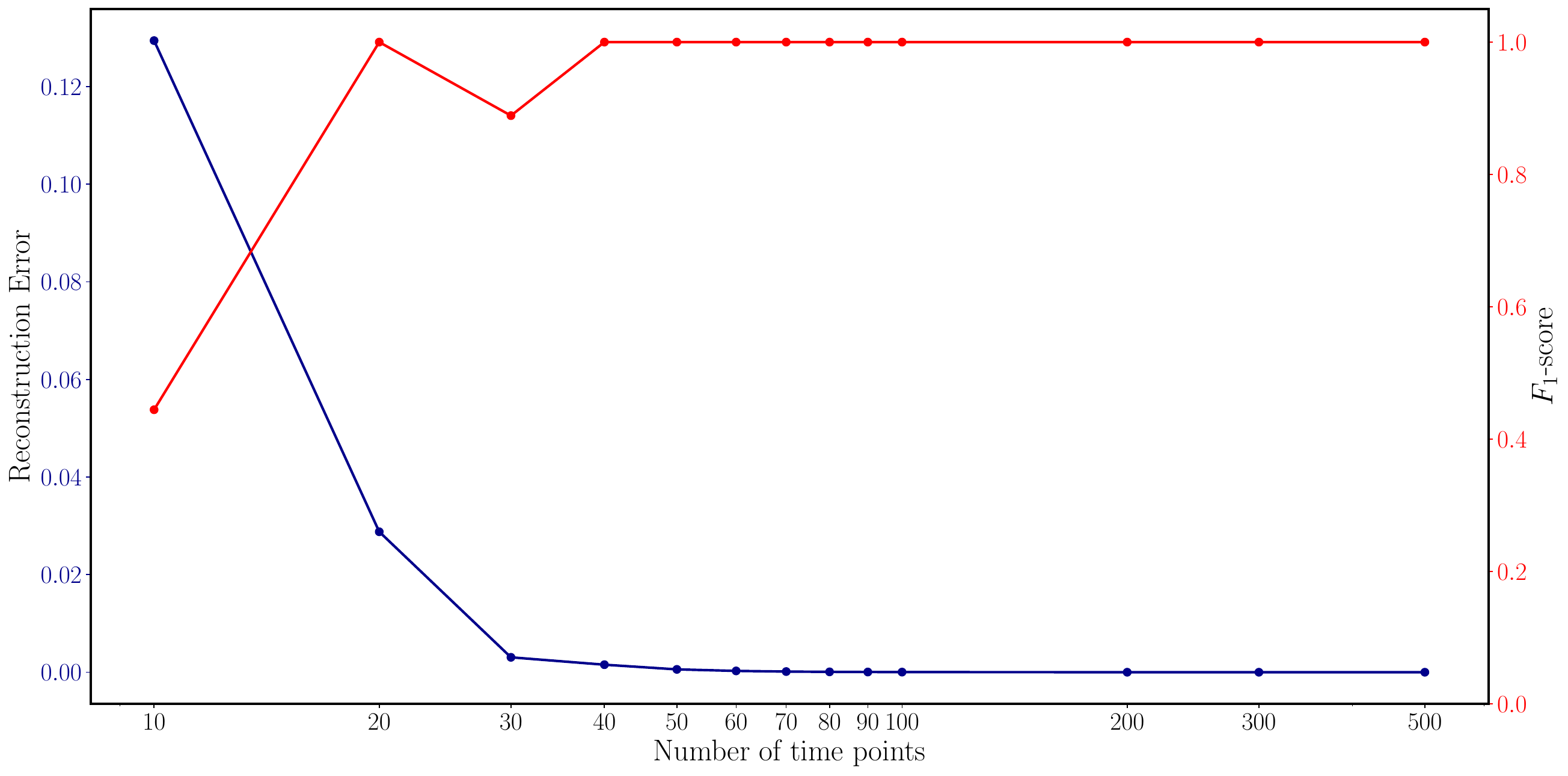}
    \caption{\textbf{Sensitivity of Reactmine to the number of time points for the chain CRN}
    with $\delta_{\text{max}}=3,\alpha=0.2,\beta=7,\gamma=5$.
    The time horizon of the simulation is $T=10$.
    The quadratic loss is reported in darkblue, the $F_1$-score in red.
    The x-axis is in log scale.}
    \label{fig:sensitimechain}
\end{figure}

\end{document}